\documentclass[aps,pra,twocolumn,groupedaddress,superscriptaddress,bibnotes,amsfonts,longbibliography,citeautoscript,a4paper]{revtex4-2}
\usepackage{latexsym}
\usepackage{amsmath}
\usepackage{amssymb}
\usepackage{graphicx}
\usepackage[T1]{fontenc}
\usepackage[open]{bookmark}
\usepackage{hyperref}
\usepackage{orcidlink}
\hypersetup{colorlinks=true,allcolors=blue}
\usepackage{xcolor} 
\usepackage{csquotes}
\usepackage{bm}
\newcommand{\vecnabla}{\bm{\nabla}}
\providecommand{\vect}[1]{{\mathbf{#1}}}
\DeclareRobustCommand{\uvec}[1]{{
  \ifcsname uvec#1\endcsname
     \csname uvec#1\endcsname
   \else
    \bm{\hat{\mathbf{#1}}}%
   \fi
}}
\newcommand{\LL}{\mathrm{L}^2}

\begin{document}

\title{
Electron beams traversing spherical nanoparticles: analytic and numerical treatment}

\author{P.~Elli~Stamatopoulou\,\orcidlink{0000-0001-9121-911X}}
\email{elli@mci.sdu.dk}
\affiliation{POLIMA---Center for Polariton-driven Light-Matter Interactions, University of Southern Denmark, 5230 Odense M, Denmark}
\author{Wenhua~Zhao$^{\ast}$\,\orcidlink{0009-0004-5721-607X}}
\affiliation{Humboldt-Universit\"at zu Berlin, Institut f\"ur Physik, AG Theoretische Optik and Photonik, 12489 Berlin, Germany}
\affiliation{Max-Born-Institut, 12489 Berlin, Germany}
\author{\'Alvaro~Rodr\'iguez~Echarri\,\orcidlink{0000-0003-4634-985X}}
\affiliation{Max-Born-Institut, 12489 Berlin, Germany}
\author{N.~Asger~Mortensen\,\orcidlink{0000-0001-7936-6264}}
\affiliation{POLIMA---Center for Polariton-driven Light-Matter Interactions, University of Southern Denmark, 5230 Odense M, Denmark}
\affiliation{Danish Institute for Advanced Study, University of Southern Denmark, 5230 Odense M, Denmark}
\author{Kurt~Busch\,\orcidlink{0000-0003-0076-8522}}
\email{kurt.busch@physik.hu-berlin.de}
\affiliation{Humboldt-Universit\"at zu Berlin, Institut f\"ur Physik, AG Theoretische Optik and Photonik, 12489 Berlin, Germany}
\affiliation{Max-Born-Institut, 12489 Berlin, Germany}
\author{Christos~Tserkezis\,\orcidlink{0000-0002-2075-9036}}
\affiliation{POLIMA---Center for Polariton-driven Light-Matter Interactions, University of Southern Denmark, 5230 Odense M, Denmark}
\author{Christian~Wolff\,\orcidlink{0000-0002-5759-6779}}
\email{cwo@mci.sdu.dk}
\affiliation{POLIMA---Center for Polariton-driven Light-Matter Interactions, University of Southern Denmark, 5230 Odense M, Denmark}

\date{\today}

\begin{abstract}

We present an analytic, Mie theory-based solution for the energy-loss and
the photon-emission probabilities in the interaction of spherical nanoparticles with electrons passing nearby and through them, in both cathodoluminescence  and electron energy-loss spectroscopies. In particular, we focus on the case of penetrating electron 
trajectories, for which the complete fully electrodynamic and relativistic
formalism has not been reported as yet.
We exhibit the efficiency of this
method in describing collective excitations in matter through calculations for a dispersive and lossy system, namely a sphere 
described by a Drude permittivity. Subsequently, we use the analytic solution to corroborate
the implementation of electron-beam sources in a state-of-the-art numerical method for problems in electrodynamics, the discontinuous Galerkin time-domain (DGTD) method. We show that the two approaches produce spectra in good mutual agreement,
and demonstrate the versatility of DGTD via simulations of spherical nanoparticles 
characterized by surface roughness. The possibility of simultaneously
employing both kinds of calculations (analytic and numerical) facilitates a
better understanding of the rich optical response of nanophotonic architectures
excited by fast electron beams. 

\end{abstract}

\maketitle

\section{Introduction}
\label{sec:introduction}

In recent decades, electron-beam spectroscopy has emerged as a revolutionary tool
for the optical characterization of materials. Swift electrons 
passing in close proximity or through a specimen undergo energy loss owing to 
energy transfer to the optical modes sustained in the 
material~\cite{GarciadeAbajo:2010rmp}. From localized and propagating surface 
plasmons in metallic 
structures~\cite{watanabe:1956,ritchie:1957,Yamamoto:2001,Vesseur:2007}, to Mie
resonances in dielectric resonators~\cite{coenen:2013,matsukata:2019,Fiedler:2022}
and phonon polaritons in polar crystals~\cite{lagos:2017,hage:2019,maciel:2020},
electron-beam spectroscopy has proven quintessential for mapping collective 
excitations in a broad spectral range that spans from ultraviolet to far-infrared
frequencies.

With the diffraction limit ultimately being controlled by the de~Broglie wavelength, highly energetic electrons are excellent probes to study
the optical properties of truly nanoscale structures, with atomic spatial resolution and sub-meV energy resolution~\cite{Polman:2019,GarciadeAbajo:2021acsp}. 
In electron energy-loss spectroscopy (EELS), the sample is excited by a high-energy electron beam ($30-300$\,keV), and the energy lost to the interaction is measured  in a transmission electron microscope (TEM) setup~\cite{egerton:2011,Wu:2018}. EELS allows, thereby,
the detection of both radiative and dark modes, including longitudinal 
bulk plasmons (BPs)~\cite{batson_silcox:1983}, breathing modes~\cite{schmidt:2012}, 
or antibonding modes in nanoparticle (NP) dimers~\cite{koh:2009}. Optical excitations
in thick
samples can be imaged in cathodoluminescence (CL) spectroscopy, performed in scanning electron microscopes (SEMs) at intermediate beam energies 
($1-50$\,keV)~\cite{kociak_um176,coenen_apr4}. In CL measurements,
the signal collected is the result of far-field photon emission from the sample, originating
from the radiative decay of the excited modes. Recent advances in instrumentation
have even added temporal resolution in EEL and CL spectra, introducing the field
of ultrafast electron microscopy (UEM)~\cite{barwick:2008,barwick:2009}.

Considering the recent progress in electron spectroscopy techniques,
robust analytic and computational tools are evidently required to interpret
the plethora of experimental data. While first theoretical efforts were
performed within the non-retarded approximation for the description of
plasmons in thin films~\cite{ritchie:1957}, the theory was gradually
generalized to account for collective excitations in diverse media and
geometries~\cite{ferrel:1985,GarciaMolina:1985}, also considering
retardation effects~\cite{rivacoba:1992,zabala:1989}. Quantum 
approaches~\cite{echenique:1987,ritchie:1988} and analytic solutions
including relativistic effects for simple geometries were later 
developed~\cite{kroger:1970,GarciadeAbajo:1999prb}, allowing the
combination of high-velocity electron beams with both common and less
conventional materials, including dielectric media, polar crystals,
graphene and other two-dimensional (2D) 
materials~\cite{Fiedler:2022,maciel:2020,hage:2020,yankovich:2019}.
Most theoretical relativistic descriptions have focused on aloof electron 
trajectories, that do not penetrate the specimen, in contrast to
experimental practices, where the electron beam is typically scanned
over the entire sample area~\cite{egerton:2011}. However, aloof electron 
trajectories oftentimes fail to capture intriguing phenomena associated with bulk
properties, such as BPs and bulk phonons~\cite{koh:2009,lagos:2017},
and other sources of electron-induced photon emission, like Cherenkov
or transition 
radiation~\cite{kroger:1970,pogorzelski_yeh:1973,ginzburg:1996,yamamoto:1996prsl}.

Despite the undeniable advantage of analytic solutions in data analysis, their applicability is limited to a handful of highly symmetric geometries.
Over the years, different numerical schemes have been consolidated to complement analytic approaches in simulating
the electromagnetic properties of nanophotonic systems~\cite{Gallinet:2015,zheng_ats2}. These include
the boundary element method (BEM)~\cite{GarciadeAbajo:2002prb,hohenester_mnpbem}, the finite element method (FEM)~\cite{Pomplun:2007,Burger:2012}, or the finite-difference frequency-domain (FDFD) and finite-difference time-domain
(FDTD) methods~\cite{pereda_fdfd,yee_ieee,taflove_ieee}, many of which have been employed successfully for CL and EELS simulations~\cite{GarciadeAbajo:2002prb,hohenester:2014,das:2012,cao:2015}.
An alternative route is offered by the discontinuous Galerkin time-domain (DGTD) method, which employs the Galerkin scheme to solve Maxwell's equations in the time domain~\cite{hesthaven2007nodal,matyssek:2011,DGTD:2011,Husnik:2013,Pramassing:2021}. 
This method combines the flexible space discretization of finite elements, with the memory efficiency and the ability to include nonlinearities, characteristic of time-domain methods. As a consequence, DGTD offers great versatility in simulating objects of complex geometry and nonlinear response. 

In this work, we present and compare an analytic approach and the DGTD method for the study of spherical nanostructures excited by aloof and penetrating electron beams. 
Following the work of Garc{\'i}a de Abajo for aloof electron 
beams~\cite{GarciadeAbajo:1999prb}, we derive  analytic formulas for the energy loss and photon emission probability, generalized here to account for 
penetrating trajectories. We then validate the implementation of electron-beam excitation of nanostructures in DGTD~\cite{matyssek:2011} by comparing the EEL and CL spectra produced by the two methods for a perfectly spherical plasmonic NP
featuring localized surface plasmons (LSPs) and BPs. Finally, we apply the numerical method to study the
optical response of a NP with surface roughness, showcasing the ability
of the DGTD method to emulate scenarios aligned with realistic experimental 
conditions that involve imperfect structures~\cite{Maradudin:1975,Maradudin:2007,loth2023surface}.

\section{Methods}
\label{sec:methods}

\subsection{Analytic approach}
\label{subsec:methods_analytics}

As a first step to examine the agreement and complementarity of analytic and numerical tools, we outline the modeling of the physical system and the assumptions made in each method. As a testbed, we consider
a perfectly spherical metal NP of radius $R$ embedded in air. The NP
is characterized by a unity relative permeability ($\mu = 1$), while its relative permittivity depends on the angular frequency $\omega$ as described by the Drude model
\begin{equation}\label{Eq:Drude}
    \varepsilon (\omega) = 
    1 - \frac{\omega^2_\mathrm{p}}{\omega (\omega + i \tau^{-1})},
\end{equation}
with plasma frequency $\omega_\mathrm{p}$ and damping rate $\tau^{-1}$.

\begin{figure}[t]
    \centering
    \includegraphics[width=0.4\columnwidth]{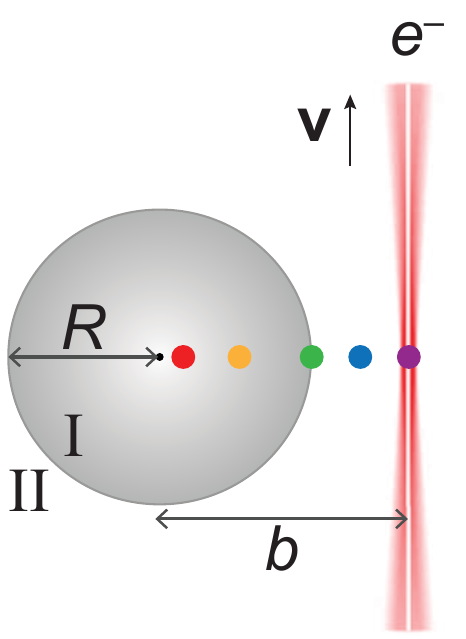}
    \caption{Schematic illustration of the geometry under study; a metallic NP of radius $R=75$\,nm, and permittivity described
by the Drude model of Eq.~\eqref{Eq:Drude}, is excited by an electron beam passing with velocity $\vect{v}$ at impact parameter $b$ with respect to its center. The color dots correspond to the selection of impact parameters examined in our study. Labels $\mathrm{I}$ and $\mathrm{II}$ indicate the region inside the NP and the surrounding medium (air), respectively.}
    \label{fig:illustration}
\end{figure}

In the analytic calculation, the electron beam is modeled as a single elementary point charge $-e$, with kinetic energy on the order of keV. Since the energy transferred to plasmon resonances is typically a few eV, we invoke the non-recoil approximation~\cite{GarciadeAbajo:2010rmp}, and assume that the electron travels with constant velocity $\vect{v}$, say along the $z$-axis. Therefore, it follows a straight
trajectory $\vect{r}_{e} = \vect{r}_{0} +\vect{v}t$, where
$\vect{r}_0 = (b, \phi_{0}, z_0=-\infty)$ is its initial position in cylindrical
coordinates. The impact parameter $b$ indicates the distance between
the electron trajectory and the center of the sphere, as shown in the
schematics of Fig.~\ref{fig:illustration}. In passing, we mention that
the angular coordinate $\phi_{0}$ need not be specified, as it does not
enter the calculation, due to the symmetry of the problem.

The fast electron drives plasmon oscillations in the metal, generating  an induced electric and magnetic field, $\vect{E}_{\mathrm{ind}}$ and $\vect{H}_{\mathrm{ind}}$ respectively, eventually resulting in the emission of radiation to the environment (labeled as region $\mathrm{II}$ in Fig.~\ref{fig:illustration}). Following the basic steps of Mie theory~\cite{mie_ap330}, one can decompose the induced electromagnetic field into transverse electric (TE) and transverse magnetic (TM) components. In region $\mathrm{II}$ the induced fields take the general form of outgoing spherical waves~\cite{jackson}
\begin{subequations}\label{far_fields}
    \begin{align}
        \vect{E}_\mathrm{ind}^\mathrm{II}(\vect{r}, \omega) &= \sum_{\ell =1}^{\infty} \sum_{m=-\ell}^{+\ell}  \Big\{ b_{\ell m}^\mathrm{II} h^+_\ell(k_0r) \vect{X}_{\ell m}(\theta , \phi)  \notag \\
        &\qquad   + \frac{i}{k_0} a_{\ell m}^\mathrm{II}  \vecnabla \times h^+_\ell(k_0r) \vect{X}_{\ell m}(\theta , \phi) \Big\},   \\
        \vect{H}_\mathrm{ind}^\mathrm{II}(\vect{r}, \omega) &= \frac{1}{Z_0}\sum_{\ell =1}^{\infty} \sum_{m=-\ell}^{+\ell}   \Big\{ a_{\ell m}^\mathrm{II} h_\ell^+(k_0r) \vect{X}_{\ell m}(\theta , \phi) \notag \\
        &\qquad   - \frac{i}{k_0} b_{\ell m}^\mathrm{II}  \vecnabla \times h_\ell^+(k_0r) \vect{X}_{\ell m}(\theta , \phi) \Big\},
    \end{align}
\end{subequations}
where $\ell$, $m$ are the angular momentum quantum numbers, and $a_{\ell m}^\mathrm{II}/b_{\ell m}^\mathrm{II}$ are the expansion coefficients of the TE/TM components corresponding to modes of electric/magnetic multipole character (see Appendix~\ref{sec:appedix_c} for the full derivation). Furthermore, in Eqs.~\eqref{far_fields} $h_\ell^+$ is the spherical Hankel function of the first kind, $\vect{X}_{\ell m}$ are the vector spherical harmonics of Eq.~\eqref{Xlm_definition}, $k_0=\omega/c$ is the wave number in free space, $c=1/\sqrt{\varepsilon_0\mu_0}$ the speed of light in vacuum, where $\varepsilon_0$ and $\mu_0$ denote the vacuum permittivity and permeability, respectively, and $Z_0 =\sqrt{\mu_0/\varepsilon_0}$ is the impedance in free space. 

The energy radiated
in the far field can be found by integrating the Poynting flux at a spherical surface of radius $r \to \infty$, in the normal direction $\uvec{r}$. Then the (CL) probability of collecting a photon of energy $\hbar\omega$ is given by
\begin{equation} \label{CL}
        \Gamma_{\mathrm{CL}} (\omega) = 
        \frac{r^{2}}{\pi \hbar\omega} 
        \int d\Omega \, 
        \mathrm{Re} \big\{ \vect{E}_{\mathrm{ind}}^\mathrm{II} (\vect{r}, \omega) \times 
        \vect{H}_{\mathrm{ind}}^{\mathrm{II}^\ast} (\vect{r}, \omega) 
        \big\} \cdot \uvec{r}, 
\end{equation}
where $d\Omega$ denotes the infinitesimal solid angle. 
By inserting Eqs.~\eqref{far_fields} into Eq.~\eqref{CL}, and evaluating the result in the far field ($k_0 r \to \infty$), we find
\begin{equation} \label{CL_decomposition}
       \Gamma_\textrm{CL} (\omega) =\frac{1}{\pi \hbar\omega Z_0 k^2_0}  \sum_{\ell =1}^{\infty} \sum_{m=-\ell}^{+\ell}  \Big\{\big| b_{\ell  m}^\textrm{II} \big|^2 + \big| a_{\ell  m}^\textrm{II} \big|^2   \Big\},  
\end{equation}
where $a_{\ell  m}^\textrm{II}$ and $b_{\ell  m}^\textrm{II}$ are given by Eqs.~\eqref{exp_coeff_corrected} in Appendix~\ref{sec:appedix_d}.

Apart from the emission of radiation, part of the energy transferred to the optical modes of the NP dissipates non-radiatively, owing to the intrinsic losses within the material. The total energy lost can be calculated by the work done by the electron against the induced field along the entire electron trajectory. Then the (EEL) probability
of the electron losing energy $\hbar\omega$ is given by
\begin{equation} \label{EELS_general}
    \Gamma_{\mathrm{EEL}} (\omega) = 
    \frac{e}{\pi\hbar\omega} 
    \int dt \, 
    \mathrm{Re} \big\{ \mathrm{e}^{-i\omega t} \,
    \vect{v} \cdot 
    \vect{E}_{\mathrm{ind}} (\vect{r}_{e}, \omega) \big\}.
\end{equation}
The integral in Eq.~\eqref{EELS_general} can be decomposed into three terms (see details in Appendix~\ref{subsec:appedix_d1})
\begin{equation} \label{EELS_decomposition}
    \Gamma_{\mathrm{EEL}} (\omega) = 
    \Gamma_{\mathrm{bulk}} (\omega) + 
    \Gamma_{\mathrm{surf}} (\omega) +  
    \Gamma_{\mathrm{Begr}} (\omega).
\end{equation}
Here, $\Gamma_{\mathrm{bulk}}$ is related to the bulk modes of the unbound medium, reduced by the Begrenzung term $\Gamma_{\mathrm{Begr}}$ that accounts for the presence of a boundary~\cite{lucas_sunjic}. The $\Gamma_{\mathrm{surf}}$ term contains the contribution from modes excited by the part of the electron trajectory lying externally to the NP, and is, thus, associated with the excitation of LSPs. The terms entering Eq.~\eqref{EELS_decomposition} are given by the following formulas
\begin{widetext}
\begin{subequations}\label{EELS_terms}
\begin{equation} \label{bulk_term}
        \Gamma_\mathrm{bulk}(\omega ) 
     = \frac{e^2 z_e}{2\pi^2\varepsilon_0 \hbar v^2} \mathrm{Im} \Bigg\{ \frac{1}{\gamma_0^2} \mathrm{ln}\left(\left[\frac{q_\mathrm{c}\gamma_0 v}{\omega}\right]^2+ 1\right) 
     - \frac{1}{\gamma^2\varepsilon} \mathrm{ln}\left(\left[\frac{q_\mathrm{c}\gamma v}{\omega}\right]^2+ 1\right) \Bigg\},
\end{equation}
\begin{align} \label{surf_term}
    \Gamma_\mathrm{surf}(\omega) 
    = \frac{e}{\pi\hbar\omega}\mathrm{Re} \sum_{\ell =1}^{\infty} \sum_{m=-\ell}^{+\ell} \Bigg\{ &\frac{K_m \left(\omega b/[v\gamma_0]\right) }{ik_0 \sqrt{\ell (\ell +1)}} \left[mb_{\ell  m}^\mathrm{II} \mathcal{M}_{\ell m}^*  - a_{\ell  m}^\mathrm{II} \frac{\mathcal{N}_{\ell m}^*}{\beta\gamma_0} \right] \notag \\
   -&\int_{-z_e}^{z_e} dz \, \frac{\mathrm{e}^{-i\omega z/v}}{\sqrt{\ell (\ell +1)}} \left[ mb_{\ell  m}^\mathrm{II} h_\ell^+(k_0r)  {Y^m_\ell} \left(\theta,0\right) - \frac{a_{\ell  m}^\mathrm{II}}{k_0b} \big\{ \mathcal{H}_{\ell m}^+(k_0z) + \mathcal{H}_{\ell m}^-(k_0z) \big\}\right] \Bigg\},    
\end{align}
and
\begin{equation}  \label{Begr_term}
    \Gamma_\mathrm{Begr} (\omega) 
      =\frac{e}{\pi\hbar\omega}\mathrm{Re} \sum_{\ell =1}^{\infty} \sum_{m=-\ell}^{+\ell} \int_{-z_e}^{z_e} dz \frac{\mathrm{e}^{-i\omega z/v}}{\sqrt{\ell (\ell +1)}}    \left[ mb_{\ell  m}^\mathrm{I} j_\ell(kr)  {Y^m_\ell} \big(\theta,0\big) 
    - \frac{a_{\ell  m}^\mathrm{I} }{kb} \big\{ \mathcal{J}_{\ell m}^+(kz) + \mathcal{J}_{\ell m}^-(kz)  \big\}\right]. 
\end{equation}
\end{subequations}
\end{widetext}
Here, $2z_e = 2\sqrt{R^2-b^2}$ is the length of the electron path inside the NP, and $\gamma = 1/[1-\varepsilon \beta^2]^{1/2}$ with $\beta=v/c$ are the Lorentz kinematic factors ($\gamma_0$ is evaluated in free space). In Eqs.~\eqref{surf_term} and \eqref{Begr_term} $K_m$ is the modified Bessel function of the second kind and $Y_\ell^m$ are the spherical harmonics. In addition, we have set $r=\sqrt{b^2+z^2}$, and $\theta =\arccos(z/r)$, while analytic expressions for coefficients $a_{\ell m}^\mathrm{I}$, $b_{\ell m}^\mathrm{I}$, $\mathcal{M}_{\ell m}$, $\mathcal{N}_{\ell m}$, $\mathcal{H}_{\ell m}^\pm$, and $\mathcal{J}_{\ell m}^\pm$ can be found in Appendices~\ref{sec:appedix_a}, \ref{sec:appedix_c} and \ref{sec:appedix_d}.

It is important to note here that the aforementioned 
decomposition of the EEL spectra introduces
a free parameter; assuming that upon losing energy $\hbar\omega$ the electron transfers a transverse (with respect to the electron trajectory) momentum $q$ to excite an optical mode, 
$q_{\mathrm{c}}$ is the maximum transverse momentum collected. In an experiment, the momentum cutoff is 
determined by the half-aperture collection angle $\varphi$ of the microscope spectrometer, as
\begin{equation}\label{Eq:cutoff}
    \hbar q_{\mathrm{c}} \approx 
    \sqrt{(m_e v \varphi)^{2} + 
    (\hbar\omega/v)^{2}},
\end{equation}
where $m_e$ is the electron mass. We may freely choose the value for 
this momentum cutoff, making sure that it aligns with the typical values for the 
collection angle in scanning TEM (STEM) setups, which are on the order of a few
mrad. Naturally, this introduces a level of arbitrariness in the EEL spectra, as different values of $q_\mathrm{c}$ lead to different peak intensities at the BP energy.

\begin{figure*}[t]
\centering
\includegraphics[width=\textwidth]{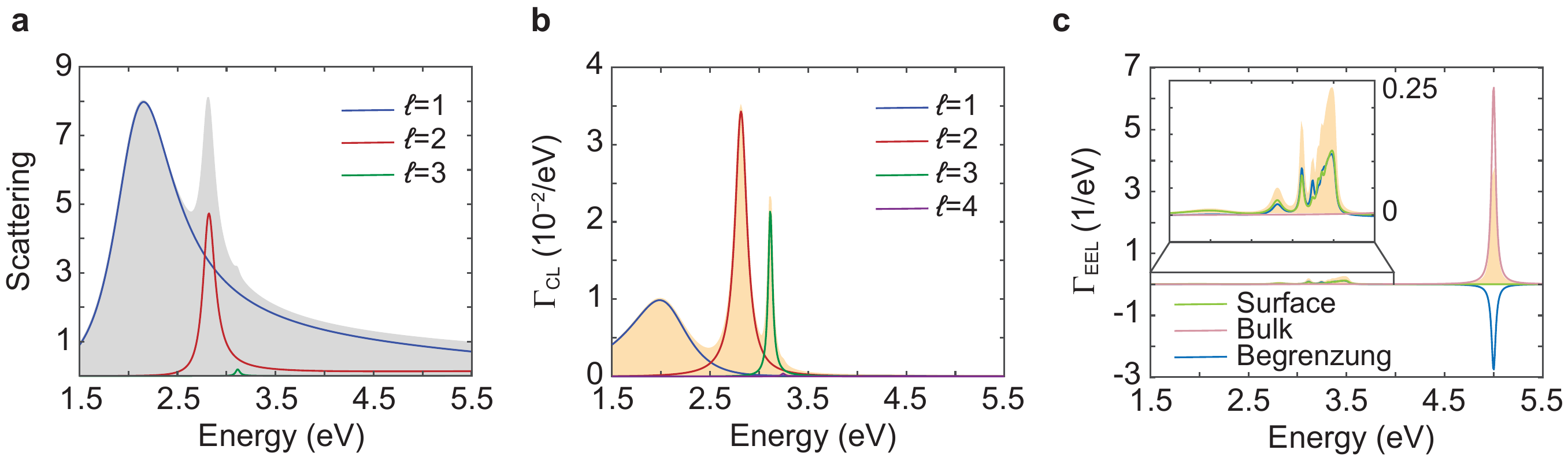}
\caption{
(a)~Scattering cross section, normalized to the geometrical 
cross section $\pi R^{2}$, of the metallic NP of Fig.~\ref{fig:illustration} under plane-wave excitation (gray shaded area). The spectrum is decomposed into the
contributions from the first three multipoles (blue, red and
green curves, for $\ell = 1, 2, 3$ respectively). In panels (b) and (c) we consider an electron traveling with velocity $v = 0.33 c$ (kinetic energy $\approx 30$\,keV) passing through the NP of Fig.~\ref{fig:illustration} at impact parameter $b = 35$\,nm. 
(b)~CL probability (yellow shaded area) and multipole decomposition
(blue, red, green, and violet curves).
(c)~EEL probability (yellow shaded area) decomposed into the surface, bulk, and Begrenzung contributions (green, pink and blue curves), with the inset focusing on the energy window $1-4$\,eV. The bulk term is given by Eq.~\eqref{bulk_term} for $q_\mathrm{c} = 0.71$\,nm$^{-1}$ and the surface and Begrenzung terms are obtained by Eqs.~\eqref{surf_term} and \eqref{Begr_term} for multipole order cutoff $\ell_\mathrm{max} = 63$.}
\label{fig:analytics}
\end{figure*}

\subsection{DGTD simulation}
\label{subsec:methods_dgtd}

We complement our analytical work with numerical simulations of the
electromagnetic problem of a NP excited by a moving Gaussian charge
distribution.
To this end, we employ the DGTD method, which combines a piecewise polynomial spatial interpolation on an unstructured tetrahedral mesh
with a Runge--Kutta time integrator to obtain a high-order accurate explicit solver for Maxwell's equations in time domain
\begin{subequations} \label{Eq:Maxwell}
\begin{align}     
    \partial_t \vect{H} (\vect{r}, t) = &   
    - \mu^{-1}_0\mu^{-1}(\vect{r}) \,  \vecnabla \times \vect{E} (\vect{r}, t),
    \\
    \partial_t \vect{E} (\vect{r}, t)=&  
    \, \varepsilon^{-1}_0\varepsilon^{-1}(\vect{r}) \, \left[ 
    \vecnabla \times\vect{H} (\vect{r}, t) - \vect{j}(\vect{r}, t)\right].
\end{align}
\end{subequations}
Here,  $\vect{j}$ is the total current density that encompasses both 
any current associated with the excitation source, as well as 
dispersive polarization currents.
The resulting method is memory-efficient compared to traditional finite elements and especially well-suited for the calculation of wide-band spectra.

One key difference between the Mie-based theory and the DGTD simulations, that can potentially lead to deviations between the two approaches, is the implementation of the excitation source. In the numerical treatment, we model the electron beam with a Gaussian charge
distribution of the form
\begin{equation}   \label{eq:Gaussiancharge}
    \rho(\vect{r}) = - \frac{ e}{\sigma_e^3 \sqrt{\pi^3} } \,\mathrm{e}^{-r^2/\sigma_e^2},
\end{equation}
with width $\sigma_e=5$\,nm. This choice essentially 
prevents numerical artifacts that would arise in a point-charge particle modeling, while also being compatible with the typical spot size in CL experiments~\cite{Fiedler:2022}. Thereby, we introduce a new spatial scale, which 
needs to be considered when it is comparable to the mesh element size and the characteristic lengths of the physical system, e.g. the radius of the NP and the impact parameter. For very large distances between the electron and the NP (i.e., $b-R\gg\sigma_e$), the source 
resembles a point charge, and we thus expect an excellent agreement 
with analytic results. However, in the opposite scenario, the finite width of the electron beam becomes important, and the corresponding fields do not accurately match those of a point charge.

\section{Results and discussion} 
\label{sec:results_discussion}

\subsection{EEL and CL spectroscopy of perfectly spherical NPs}
\label{subsec:smooth_sphere}

\begin{figure*}[t]
    \centering
    \includegraphics[width=\textwidth]{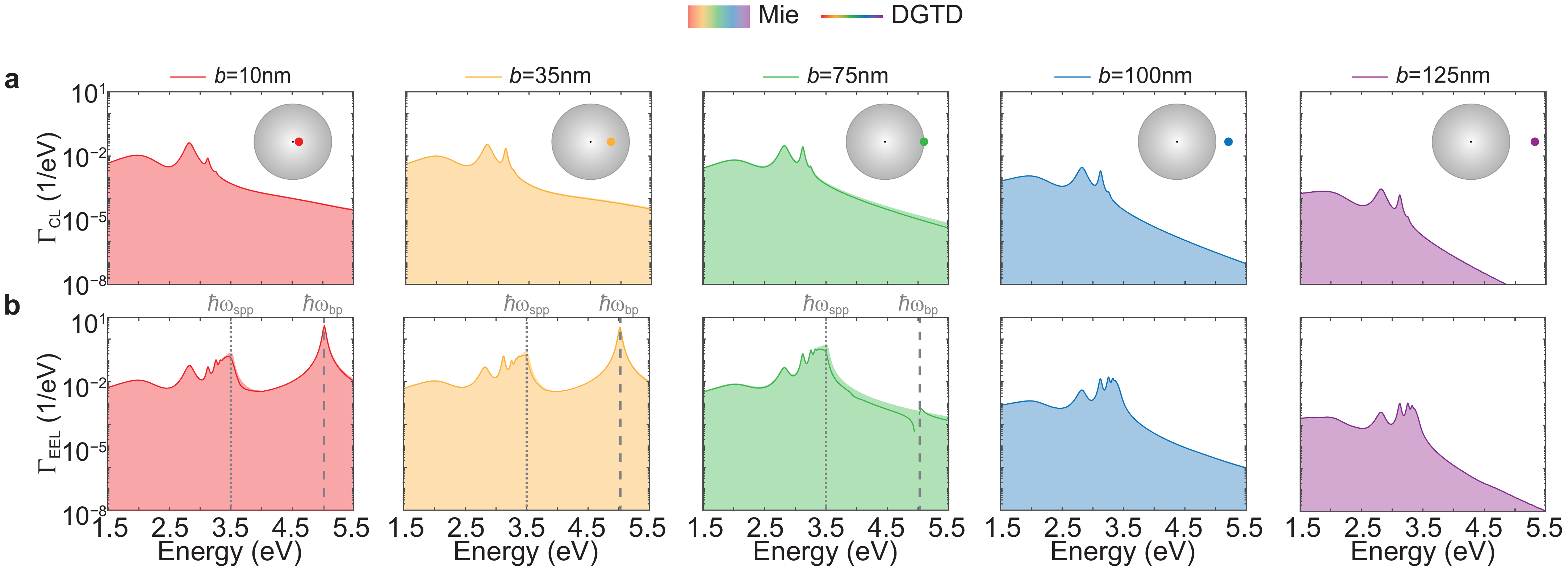}
    \caption{(a)~CL and (b)~EEL probability in the interaction between the metallic NP of Fig.~\ref{fig:illustration} and an electron traveling with velocity $v =0.33c$ (kinetic energy $\approx 30$\,keV) considering 5 different impact parameters, as denoted in the labels and illustrated in the insets of each panel. The shaded areas correspond to results of the analytic Mie-type 
    calculation, whereas the solid lines show the corresponding DGTD simulations. The vertical gray dotted and dashed lines in the three leftmost panels of (b) trace the energy of the SPP ($\hbar\omega_\mathrm{spp} \approx 3.5$\,eV) and the BP mode ($\hbar\omega_\mathrm{bp} \approx 5$\,eV), respectively. In the two leftmost panels of (b) we use $q_\mathrm{c}=0.71\,\mathrm{nm}^{-1}$, while $\ell_\mathrm{max}=63$ in all panels.}
    \label{fig:perfect_sphere_comparison}
\end{figure*}

In what follows, we analyze the response of a metallic NP excited by a fast electron beam, as predicted by both the analytic and the numerical approach. We consider a smooth sphere of radius $R = 75$\,nm with plasma energy $\hbar\omega_\mathrm{p} = 5$\,eV and damping rate $\tau^{-1}$
corresponding to an energy of $\hbar\tau^{-1} = 50$\,meV. The parameters mimic typical plasmonic metals and are chosen for the purpose of illustration, while the particular values have no consequences for our general conclusions. 
Fig.~\ref{fig:analytics} shows the scattering spectrum (panel~a) under plane-wave illumination, and the CL and EEL probability (panels~b and c, respectively) calculated for a low electron velocity $v =0.33 c$ (kinetic energy $\approx 30$\,keV)  intersecting the NP at 
$b = 35$\,nm from its center. Eq.~\eqref{CL_decomposition} suggests that the total photon emission probability can be decomposed into contributions of 
pairwise orthogonal
electric and magnetic multipoles of order $\ell$. Fig.~\ref{fig:analytics}b reveals that the CL spectrum is composed of the contributions of the
first four ($\ell = 1, 2, 3, 4$) electric-type modes, appearing at approximately $2$, $2.8$,
$3.1$, and $3.2$\,eV, associated with the excitation of LSP resonances, while
higher-order ($\ell >4$) multipoles contribute negligibly to the spectrum. In
comparison with the scattering cross section of the NP shown in Fig.~\ref{fig:analytics}a (calculated using Mie
theory~\cite{Bohren_Wiley1983}),
the CL spectrum provides very similar information. This is somewhat expected, 
since both calculations are based on the collection of far-field radiation. We 
observe, nonetheless, two notably distinct features in the CL spectrum of Fig.~\ref{fig:analytics}b. Firstly, the electron source excites more efficiently the $\ell=2$ and $3$ modes, whereas in the scattering spectrum of Fig.~\ref{fig:analytics}a the dipolar mode peak features the highest intensity. The relative peak intensities in the CL and EEL spectra depend strongly on the impact parameter, which determines the arrangement of the polarization charges in the material~\cite{batson_nl11}.
Secondly, we observe a small redshift of the dipolar ($\ell = 1$) mode in CL due to retardation, stemming from the fact that the speed
of the electron is only a fraction of the speed of light.

In Fig.~\ref{fig:analytics}c we present the EEL spectrum, decomposed as described in Eq.~\eqref{EELS_decomposition}. 
In the zoom-in area of the figure we observe the
excitation of numerous higher-order multipoles, appearing as sharp peaks at energies up to $3.5$\,eV. With increasing multipole order, the wavelength of
the corresponding mode reduces, and higher-order modes experience the curved 
surface of the NP as increasingly more flat. As a result, they accumulate at the energy
corresponding to that of a surface plasmon polariton (SPP) at a planar interface, at
$\hbar \omega_{\mathrm{spp}} = \hbar\omega_{\mathrm{p}} /\sqrt{2} \approx 3.5$\,eV.  
Above the SPP energy, the spectrum exhibits a pronounced peak at the BP energy
$\hbar\omega_{\mathrm{bp}} \approx \hbar\omega_{\mathrm{p}} = 5$\,eV, pertaining to the excitation of BPs in the volume of the NP. Since BPs are 
longitudinal modes, they do not couple to far-field radiation and, therefore, can be detected only in EELS.
At the same energy, we observe the expected negative peak related to the Begrenzung term, reducing the BP peak in the total EEL probability~\cite{lucas_sunjic}. 

Having a clear picture of the origin of all spectral features, in Figs.~\ref{fig:perfect_sphere_comparison}a and b we compare the CL and  
EEL probability, respectively, of the same metal NP, as calculated employing the analytic (Mie) and the DGTD method. We test
the agreement between the two calculations probing various impact parameters, that range
from $b = 125$\,nm to $10$\,nm, corresponding to aloof electron trajectories 
(violet and blue spectra), grazing (green), and penetrating (yellow and red). 
Overall, the DGTD method reproduces the positions of the LSP modes and the corresponding CL probabilities of the Mie calculations. In the CL spectra of Fig.~\ref{fig:perfect_sphere_comparison}a we find an excellent agreement, with a relative error [as given by Eq.~(S.11) in the SI] of around $1\%$ for aloof and penetrating electron trajectories, according to Table~\ref{tab:relative_error}. 
The highest error is acquired when the electron grazes the surface of the NP, passing exactly at $b=R$ (middle panel in Fig.~\ref{fig:perfect_sphere_comparison}a). This point reflects an important limit in the 
capabilities of the DGTD method; in the grazing trajectory, due to the finite width of the electron beam, half of the Gaussian charge density distribution lies inside the NP, while the other half lies outside, leading to numerical inconsistencies. One may avoid this point, which is inevitably difficult
to resolve, by slightly adjusting the impact parameter by half the Gaussian
width. 

Regarding compatibility of the EEL spectra, the two rightmost panels in Fig.~\ref{fig:perfect_sphere_comparison}b reveal an excellent agreement for aloof
electron trajectories. We consistently find a higher error for grazing trajectories (middle panel in Fig.~\ref{fig:perfect_sphere_comparison}b); here, the EEL spectrum calculated with DGTD exhibits a numerical artifact at the BP energy (gray dashed line), resulting once again from the fact that just a fraction of the electron charge density distribution penetrates the NP, exciting only partially the bulk mode.
A substantial deviation between the two methods is found in the EEL
spectra for both grazing and penetrating electron trajectories (three leftmost panels in Fig.~\ref{fig:perfect_sphere_comparison}b) between the SPP and the BP energy, denoted by the gray dotted and dashed lines, respectively. The disagreement is, naturally, reflected in the large relative errors presented in Table~\ref{tab:relative_error} accordingly. At these impact parameters, the condition $b-R \gg \sigma_e$ is not fulfilled, hence the field of the electron deviates considerably from that of a point charge. Moreover, the beam width $\sigma_e$ becomes important compared to the mesh element size, since the surface of the NP is more finely discretized than the surrounding medium.
Finally, there exists an additional source of error in the evaluation of the BP contribution, that stems from the rather arbitrary choice of the transverse momentum cutoff $q_\mathrm{c}$ in the analytic approach. In contrast, in the DGTD implementation there is a respective internal limit, associated with $\sigma_e$. 
As a result, we consistently find higher relative errors in EELS in comparison with CL, and for grazing and penetrating electron trajectories as compared to aloof. This is evident for both low-energy electron beams as in Fig~\ref{fig:perfect_sphere_comparison}, as well as higher energies that are more realistic for EEL measurements. 

\begin{table}[t]
\caption{\label{tab:relative_error}
Relative error between the analytic and the DGTD calculations of $\Gamma_\mathrm{CL}$ and $\Gamma_\mathrm{EEL}$ for varying $b$.}
\begin{ruledtabular}
\begin{tabular}{ccc}
 $b$ (nm) &$\Gamma_\mathrm{CL}$ (\%) & $\Gamma_\mathrm{EEL}$ (\%)\\
\hline
10& 1.26 & 9.75 \\
35& 1.06 & 6.17  \\
75& 8.23 & 28.87 \\
100& 1.06 & 1.02 \\
125& 1.07 &1.93 \\
\end{tabular}
\end{ruledtabular}
\end{table}

Admittedly, the accumulation point of high-order multipoles at $\hbar\omega_\mathrm{spp}$ is hard to resolve in both methods. On the one hand,
the analytic Mie calculation assumes a moving point charge,
which can, in principle, excite an infinite number of multipoles, resulting in a sharp
high-intensity peak at energy $\hbar\omega_{\mathrm{spp}}$. On the other hand, the finite mesh
size and beam width implemented in DGTD imposes a limitation to the number of
multipoles that can be resolved for a given discretization, since high order multipoles associated with field variation shorter than the mesh element size at the surface cannot be captured without the use of very high order polynomials. In EELS and CL experiments, there exists an analogous limitation, associated with the finite width of the electron beam employed, as well as the geometric imperfections of the NP. The versatility of DGTD allows us not only to adjust the electron beam width according to the experimental setup, but also to mimic NPs with surface roughness, as we discuss in Section~\ref{subsec:rough_sphere}. In the analytic approach, the smearing of higher-order modes and the overall quenching of the sharp peak at the accumulation point can too be reproduced, once we consider the nonlocal response of the material; this can be done particularly easily within Mie theory~\cite{Christensen:2014,zouros:2020,gonccalves:2023}. Nonlocal effects manifest as increased damping and uneven energy shifts of high-order modes, and, therefore, lead to the suppression of the individual modes, as well as the reduction of their overlap at $\hbar\omega_{\mathrm{spp}}$~\cite{moeferdt:2017}. 

\begin{figure}[t]
    \centering
    \includegraphics[width=\columnwidth]{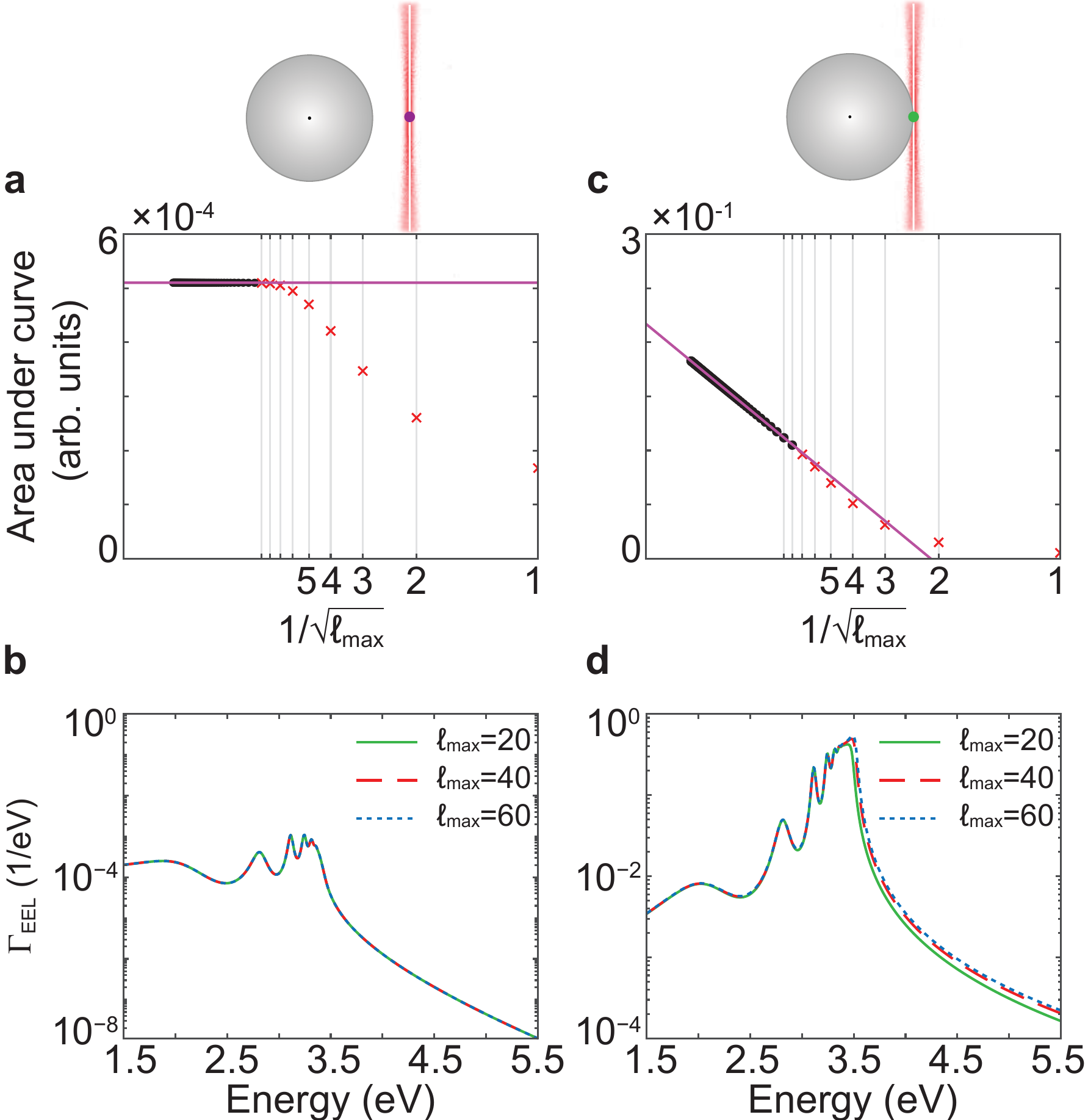}
    \caption{
    (a,~c)~Area under the curve of the EEL spectra of Fig.~\ref{fig:perfect_sphere_comparison}b, (a)~for the aloof electron trajectory at $b=125$\,nm, (c)~for the grazing trajectory at $b=75$\,nm, showcasing the convergence of the EEL probability for increasing values of the multipole cutoff $\ell_\mathrm{max}$, plotted versus $1/\sqrt{\ell_\mathrm{max}}$. In both panels, the magenta line is fitted to the data points marked as black bullets, while the red crosses represent data points excluded from the fitting. The vertical gray lines serve as guides to the eye for the position of $\ell_\mathrm{max} =1,2,\ldots 9$. (b,~d)~EEL probability calculated for impact parameters (b)~$b=125$\,nm, and (d)~$b=75$\,nm, and for selected values of $\ell_\mathrm{max}$, as denoted in the labels.}
    \label{fig:convergence_analytics1}
\end{figure}

\begin{figure}[t]
    \centering
    \includegraphics[width=\columnwidth]{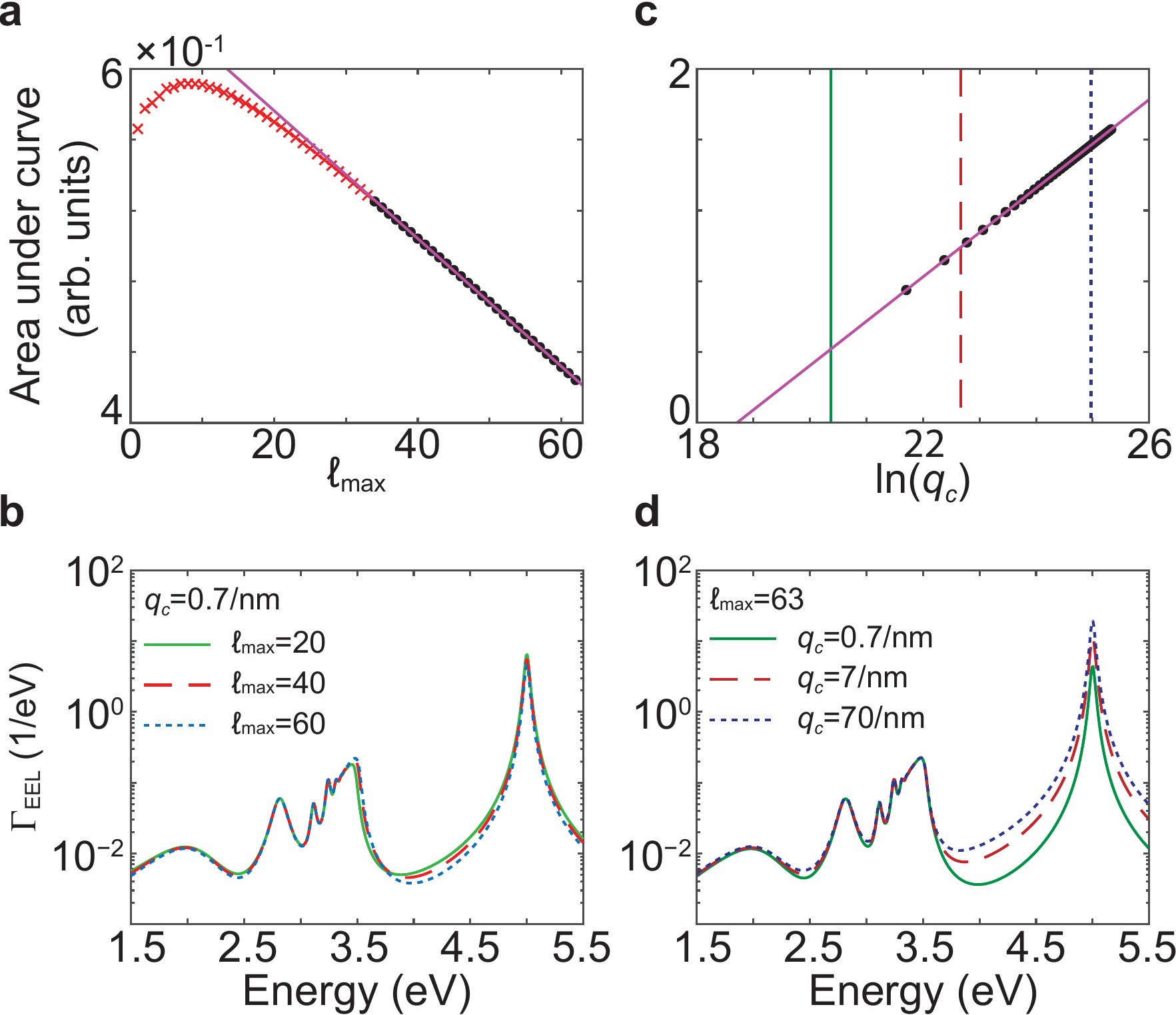}
        \caption{(a,~c)~Area under the curve of the EEL spectrum of the leftmost panel in Fig.~\ref{fig:perfect_sphere_comparison}b ($b=10$\,nm), showcasing the divergence of the EEL probability for increasing values of (a)~the multipole cutoff $\ell_\mathrm{max}$, and (c)~the transverse momentum cutoff $q_\mathrm{c}$ plotted versus $\mathrm{ln} (q_\mathrm{c})$. The data follow a linear divergence trend with respect to their corresponding horizontal axes. In both panels,
        the magenta line is fitted to the data points marked as black
        bullets, while the red crosses represent data points excluded
        from the fitting. 
        (b,~d)~EEL probability calculated at $b=10$\,nm for selected values of (b)~$\ell_\mathrm{max}$, and (d)~$q_\mathrm{c}$. In panels (a,~b) we scan over $\ell_\mathrm{max}$, while keeping a fixed value $q_\mathrm{c} = 0.7$\,nm$^{-1}$, whereas in (c,~d)~we scan over $q_\mathrm{c}$ for a fixed value $\ell_\mathrm{max} = 63$, as denoted in the labels. In panel (c) the vertical lines correspond to the selected $q_\mathrm{c}$ values presented in (d), following the same color coding.}
    \label{fig:convergence_analytics2}
\end{figure}

The difficulty in resolving the high-order multipoles even in the analytic 
calculation is clear in the convergence study presented in 
Figs.~\ref{fig:convergence_analytics1} and \ref{fig:convergence_analytics2}. Due to multipole orthogonality, the sole contribution of increasing $\ell_\mathrm{max}$  is to amplify the peak at $\hbar\omega_\mathrm{spp}$. This suggests monitoring the total EEL spectrum 
integrated over energy (area under the curve) as a proxy for convergence.
For aloof trajectories, the asymptotics of the Hankel functions suggest exponential
convergence, which is in agreement with Figs.~\ref{fig:convergence_analytics1}a and b;
the EEL probability converges at $\ell_\mathrm{max}=10$.
In contrast, Fig.~\ref{fig:convergence_analytics1}c shows that in the case of the 
grazing trajectory the convergence order breaks down to a square root law.
By extrapolation (magenta line) we can conclude that even for $\ell_\mathrm{max}=63$ the
analytic result is still converged only up to about $15\%$.
Fig.~\ref{fig:convergence_analytics1}d corroborates that the area missing
from the converged value corresponds to the higher-order modes piling up
at the SPP energy. 
Finally, we note that Fig.~\ref{fig:convergence_analytics1}a exhibits the same square
root convergence before the curve flattens off.
The exact value of $\ell_\mathrm{max}$ where this transition happens increases as the impact
parameter approaches $R$.

For the penetrating trajectories shown in Fig.~\ref{fig:convergence_analytics2} we find signs of the same slow convergence around the SPP accumulation point.
This is masked in our convergence plots by the fact that the BP peak diverges.
The source of the divergence lies in the decomposition of the EEL probability into two competing contributions stemming from the bulk and the Begrenzung terms. As illustrated in Figs.~\ref{fig:convergence_analytics2}a and c, the two
terms produce divergences of opposite sign; the
negative Begrenzung term diverges linearly for increasing multipole order $\ell_\mathrm{max}$, whereas the positive bulk
term diverges logarithmically for increasing momentum cutoff $q_\mathrm{c}$ [see Eqs.~\eqref{bulk_term} and \eqref{Begr_term}].
As computational resources do not allow driving $\ell_\mathrm{max}$ and $q_\mathrm{c}$ to
infinity, once one of the two parameters is truncated to a certain cutoff value,
the other has to be adjusted accordingly. Figs.~\ref{fig:convergence_analytics2}b
and d show that the different values of $\ell_\mathrm{max}$ and $q_{\mathrm{c}}$, respectively, affect the spectra for energies above the SPP mode. 

\subsection{EELS of NPs with surface roughness} \label{subsec:rough_sphere}

The synthesis of metallic NPs, as the ones studied in the present work, is routinely done with colloidal chemistry, for a large variety of materials and NP shapes~\cite{link1999size,grzelczak_nn4}. However, despite being able to accurately control the NP size, assuring a smooth surface is rather challenging. Typically the structures exhibit protuberances on the surface, which can be responsible for symmetry breaking~\cite{nappa2005electric,salomon2013size}, hot spots in dimers~\cite{zheng2015tailoring,yoon2019surface} and picocavities~\cite{benz2016single,urbieta_acsn12}, or energy shifts of the LSPs~\cite{rodriguez2009effect}. 
Within the DGTD method, surface texture can be easily implemented on top of the perfectly spherical mesh and incorporated in the numerical calculations. Here, we follow the prescription presented in Ref.~\cite{loth2023surface} to implement the desired roughness. Given an initial smooth sphere of radius $R$, we introduce deviations from its nominal value in both the normal and the lateral direction, which are defined by the root-mean squared (rms) roughness value and the correlation length $l$, respectively. The rms parameter is related to the local variations of the radius that follow a Gaussian white noise distribution, while $l$ describes the average distance between neighboring bumps on the surface.

\begin{figure}[t]
    \centering
    \includegraphics[width=0.8\columnwidth]{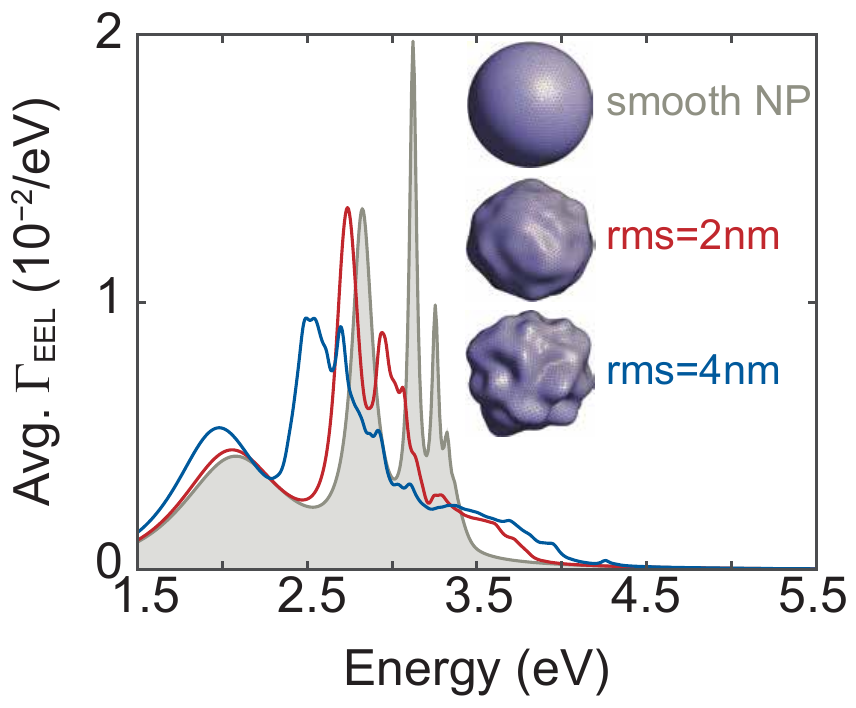}
        \caption{Average EEL probability in the interaction between a spherical NP featuring surface roughness and an electron beam passing with velocity $v = 0.7 c$ (kinetic energy $\approx 200$\,keV) at distance $b=100$\,nm. The solid lines correspond to NPs, whose shapes deviate from that of a perfectly smooth sphere of radius $R=75$\,nm (gray shaded area) by root-mean square roughness values $\mathrm{rms}=2$\,nm (dark red curve) and $\mathrm{rms}=4$\,nm (dark blue curve). The average EEL probability corresponds to the average values obtained with DGTD for $6$ different rough meshes characterized by the same $\mathrm{rms}$. 
        }
    \label{fig:rough_sphere}
\end{figure}

In Fig.~\ref{fig:rough_sphere} we explore the effect of surface roughness on a spherical NP, described by the same Drude permittivity as in the previous section, now excited by an electron beam traveling with velocity $v=0.7c$ (kinetic energy $\approx 200$\,keV) at distance $b=100$\,nm from its center. We probe meshes of two degrees of surface roughness on top of the NP of nominal radius $R=75$\,nm, namely $\mathrm{rms}=2$\,nm and $4$\,nm, while the correlation length is fixed at $l=10$\,nm in both cases. 
Since the breaking of the spherical symmetry introduces a dependence on the electron propagation direction, and on the mesh morphology, in Fig.~\ref{fig:rough_sphere} we plot the \emph{average} EEL probability, corresponding to the average values obtained for $6$ different meshes. 
The resulting spectra for $\mathrm{rms}=2$\,nm and $4$\,nm (dark red and blue curves, respectively) deviate notably from that of a smooth sphere (gray shaded area). Firstly, we observe an increasing redshift of the spectra with increasing degree of roughness, in agreement with experimental observations of corrugated plasmonic NPs~\cite{rodriguez2009effect}. The energy shift is most evident for the  dipolar mode at around $2$\,eV, and is the result of the area increase of the rough NP as compared to the smooth one. Evaluation of this area from the mesh parameters yields an effective radius of $R_\mathrm{eff}=76.2$\,nm for $\mathrm{rms}=2$\,nm, and $R_\mathrm{eff}=79.6$\,nm for $\mathrm{rms}=4$\,nm. Indeed the EEL spectra of smooth spheres of said effective radii reproduce accurately the position of the dipolar mode (see Appendix~\ref{sec:appendix_e}).

In addition to the redshift, the spectra of the corrugated NPs feature a large number of low-intensity peaks. These new spectral features are the result of two factors. Due to the breaking of the spherical symmetry, the prior degenerate modes associated with the same angular momentum $\ell$ but different $m$ number, now exhibit a small energy difference. As a result of the lift of the degeneracy, the sharp peaks observed in the spectrum of the smooth NP are suppressed and, depending on the size of this energy difference with respect to the linewidth of the degenerate mode, they are either split or broadened. 
Moreover, additional spectral features may arise from hot spots, namely protuberances of large curvature that strongly enhance and confine the incident field~\cite{urbieta_acsn12}. It is important to note here that, as the features become increasingly smaller, a rigorous description of the system requires the implementation of nonlocal effects in the method~\cite{stamatopoulou:2022,wegner_prb107}, which effectively introduces a cutoff in the contribution of large-wavevector components~\cite{wiener_nl12,Mortensen2021review}.

\section{Conclusions}

We have presented an analytic and a numerical method for the study of spherical structures excited by fast electron beams. Based on Mie theory, we have derived formulas for the calculation of the EEL and CL probability that are valid for both aloof and penetrating electron beams, as is typically the practice in EEL and CL measurements. Focusing on the plasmon oscillations of a metallic NP as a testbed, we compared the analytic theory with numerical simulations performed using the DGTD method, and found excellent agreement. We discussed the applicability and limitations of each method, particularly for grazing trajectories and at energies near the surface- and bulk-plasmon resonances. Finally, we showcased the flexibility of the DGTD method by studying a NP with different degrees of surface corrugation, which can lead to resonance shifts and splittings due to the lifting of mode degeneracy. We thus believe that both methods are essential and complementary for exploring collective optical excitations in matter, and for interpreting experimental observations. 
\vspace{0.5cm}

\section*{Acknowledgments}

We thank  F. Intravaia, C. Maciel-Escudero, and B. Beverungen for stimulating discussions.
K.~B. acknowledges funding by the German Research Foundation (DFG) in the framework of the Collaborative Research Center 1375 “Nonlinear Optics down to Atomic Scales (NOA)” (Project No.~398816777).
N.~A.~M. is a VILLUM Investigator supported by VILLUM Fonden (grant No.~16498).
The Center for Polariton-driven Light-Matter Interactions (POLIMA) is funded by the Danish National Research Foundation (Project No.~DNRF165).

\section*{Author contribution}

P.~E.~S and W.~Z. contributed equally to this work. P.~E.~S performed the  analytic study and W.~Z. the DGTD simulations. All authors participated in the discussion of the results and the writing of the manuscript. 

\appendix

\section{Electromagnetic field of a fast electron in free space} 
\label{sec:appedix_a}

The electromagnetic field of a point charge traveling in free space can be found in most standard electrodynamics textbooks, see for instance Ref.~\cite{jackson}. Here, we follow the work of Garc\'{i}a de Abajo~\cite{GarciadeAbajo:1999prb,GarciadeAbajo:2010rmp}, in which the field is decomposed into TE and TM waves.
Without loss of generality, we assume that the electron travels along the $z$-axis ($\vect{v} =v \uvec{z}$), in a straight trajectory $\vect{r}_e (t)= \vect{r}_0 + \vect{v}t$, where $\vect{r}_0 = (b, \phi_0, z_0=-\infty)$  denotes its initial position in cylindrical coordinates. The electron is modeled as a point particle of charge $-e$, and charge density $\rho (\vect{r}, t) = -e\delta[\vect{r} -\vect{r}_e(t)]$. 

By introducing the angular momentum operator $\vect{L} =  -i\vect{r} \times \vecnabla $, the electric field produced by the electron can be decomposed into TE and TM waves in the frequency domain, as~\cite{muller_jmp28} 
\begin{equation}
    \vect{E} (\vect{r}, \omega)=  \vect{L} \psi^{\mathrm{M}}(\vect{r}, \omega)+\frac{i}{k_0}\vecnabla \times \vect{L} \psi^{\mathrm{E}}(\vect{r}, \omega),
\end{equation}
where $k_0$ is the wave number in free space, and the scalar functions $\psi^{\mathrm{M/E}}$ satisfy the expressions
\begin{subequations}
    \begin{gather}
        \LL\psi^{\mathrm{M}}(\vect{r}, \omega) = \vect{L} \cdot \vect{E}(\vect{r}, \omega),\\
        -\frac{i}{k_0}\LL\nabla^2\psi^{\mathrm{E}} (\vect{r}, \omega)= (\vect{L} \times \vecnabla )\cdot \vect{E}(\vect{r}, \omega).
    \end{gather}
\end{subequations}
In the absence of charges and currents, the electric field satisfies the vector wave equation $\left( \nabla^2 + k_0^2 \right) \vect{E} (\vect{r}, \omega) = 0$. As a result,
the scalar functions satisfy the scalar wave equation 
\begin{equation}\label{scalar_wave_eq}
    \left( \nabla^2 + k_0^2 \right) \psi^{\mathrm{M/E}} (\vect{r}, \omega)=0,
\end{equation}
whose Green's function in spherical coordinates is given by~\cite{jackson}
\begin{align} \label{Greens_spherical}
    &\mathcal{G} (\vect{r} -\vect{r}') = \frac{\mathrm{e}^{ik_0|\vect{r}-\vect{r}'|}}{4\pi |\vect{r}-\vect{r}'|}  \notag \\
    &= ik_0 \sum_{\ell=0}^\infty  j_ \ell (k_0r_<) h_\ell^+(k_0r_>)   \sum_{m=-\ell}^{\ell}  Y_\ell^m (\theta , \phi){Y_\ell^m}^* (\theta' , \phi'),
\end{align}
where $ Y_\ell^m$ are the scalar spherical harmonics. In Eq.~\eqref{Greens_spherical}, we have introduced the angular momentum quantum numbers $\ell$ and $m$, and $r_{</>}$ stands for the smaller/greater between two points $r$ and $r'$. 
Eq.~\eqref{scalar_wave_eq} admits standing and outgoing spherical waves as solutions, represented by the spherical Bessel function $j_\ell$, and the spherical Hankel function of the first kind $h^+_\ell$, respectively. 
Thus, we expand the electric field at points $r < |\vect{r}_e|$ as follows
\begin{align} \label{E_general_electron}
    \vect{E}(\vect{r}, \omega) &= \sum_{\ell =1}^{\infty} \sum_{m=-\ell}^{+\ell}  \Big\{ b_{\ell m}^0 j_\ell(k_0r) \vect{X}_{\ell m}(\theta , \phi) \notag \\
    &\qquad\quad + \frac{i}{k_0} a_{\ell m}^0  \vecnabla \times j_\ell(k_0r) \vect{X}_{\ell m}(\theta , \phi)  \Big\},
    \end{align}
where $\vect{X}_{\ell m}(\theta , \phi) $ are the vector spherical harmonics, defined as
\begin{equation} \label{Xlm_definition}
    \vect{X}_{\ell m} (\theta , \phi) = \frac{1}{\sqrt{\ell (\ell +1)}} \vect{L} Y_\ell^m (\theta , \phi), 
\end{equation}
with $\vect{X}_{00}=0$. The expansion coefficients in Eq.~\eqref{E_general_electron} are given by
\begin{subequations} \label{exp_coeff_aloof}
    \begin{equation}\label{exp_coeff_b_aloof}
    b_{\ell m}^0 =  -\frac{ik_0^2e}{\varepsilon_0\omega} \frac{m \mathcal{M}_{\ell m}}{\sqrt{\ell (\ell +1)}}  K_m \left( \frac{\omega b}{v\gamma_0}\right) \mathrm{e}^{-im\phi_0},
\end{equation}
    and
    \begin{equation} \label{exp_coeff_a_aloof}
        a_{\ell m}^0 = \frac{ik_0^2e}{\varepsilon_0\omega} \frac{1}{\beta\gamma_0} \frac{\mathcal{N}_{\ell m}}{\sqrt{\ell (\ell +1)}} K_m \left( \frac{\omega b}{v\gamma_0}\right) \mathrm{e}^{-im\phi_0},
    \end{equation}
\end{subequations}
with $\gamma_0 = 1/\sqrt{1-\beta^2}$ and $\beta=v/c$ being the Lorentz kinematic factors, and $K_m$ the modified Bessel function of the second kind. In Eqs.~\eqref{exp_coeff_aloof} we have set
\begin{equation}
    \mathcal{M}_{\ell m} = i^{\ell + m} \sqrt{\frac{2\ell+1}{\pi}\frac{(\ell-m)!}{(\ell+m)!}} \frac{(2m-1)!!}{(\beta\gamma_0)^m} G_{\ell - m}^{m+1/2} \bigg( \frac{1}{\beta}\bigg), 
\end{equation}
which holds for $m \geq 0$, while $\mathcal{M}_{\ell -m} = (-1)^m \mathcal{M}_{\ell m}$, and $G_{\ell - m}^{m+1/2}$ is the Gegenbauer polynomial.
Furthermore, we have set
\begin{equation}
    \mathcal{N}_{\ell m} = c_{\ell}^m \mathcal{M}_{\ell \, m+1} -c_{\ell}^{-m} \mathcal{M}_{\ell \, m-1},
\end{equation}       
with
\begin{equation}
    c_{\ell}^{m} = \frac{1}{2}\sqrt{(\ell -m )(\ell +m +1)}.     
\end{equation}

\section{Electromagnetic field of a fast electron in the bulk of an unbound medium}
\label{sec:appedix_b}

Consider the system illustrated in Fig.~\ref{fig:SI_illustration}b, depicting the electron traversing the distance $2z_e$ inside a medium of relative permittivity $\varepsilon$ and relative permeability $\mu=1$. In principle the field generated by the electron can be found via Eq.~\eqref{E_general_electron}, with modified optical parameters. However, this expansion of the field in spherical waves is based on the Ansatz that the field is composed of a TE and a TM field component. While this is a reasonable assumption for fields in free space, inside a polarizable medium longitudinal modes can also appear. In fact, the metallic NP studied here
sustains longitudinal BPs, which cannot be described by Eq.~\eqref{E_general_electron}. It is, therefore, more convenient to avoid the field expansion in spherical waves and follow a different route, working with fields in the Cartesian coordinate system.

The charge density of the electron is given by $\rho (\vect{r}, t) = -e\delta(x)\delta(y)\delta(z-vt)$, or, by performing a Fourier transform with respect to $x,y$ and $t$, $\rho (q,z, \omega) = -(e/v) \mathrm{e}^{i\omega z/v}$. After the same Fourier transform, the vector wave equation for the field generated by the electron takes the form
\begin{align}
    \left(-q^2 +k^2 +\partial_z^2 \right) \vect{E} (q, z, \omega) 
    = -\frac{ie}{\varepsilon\varepsilon_0 v} \left(q\uvec{R} + \frac{\omega}{v\gamma^2}\uvec{z} \right) \mathrm{e}^{i\omega z/v},
    \label{helmholtz_FT}
\end{align}
where $k=\sqrt{\varepsilon}\omega/c$ is the wave number inside the infinite medium, $q^2 = k_x^2+k_y^2$, $\uvec{R} = \uvec{x}, \uvec{y}$, and $\gamma= 1/\sqrt{1-\varepsilon\beta^2}$ is the Lorentz factor evaluated in the medium. One can then readily derive the $z$ component of the field, as
\begin{equation} \label{eq:field0_z_comp}
    {E}_z (q, z, \omega) = \frac{ie}{\varepsilon\varepsilon_0} \frac{\omega}{v^2\gamma^2} \frac{\mathrm{e}^{i\omega z/v}}{q^2-k^2+(\omega/v)^2}.
\end{equation}

\section{Electromagnetic field of a nanosphere excited by a fast electron}
\label{sec:appedix_c}

Here, we derive the electromagnetic field generated by a swift electron passing through a spherical structure, taking relativistic and retardation effects into account. Even though in the present work we focus on metallic NPs that sustain LSP and BP modes, the expressions derived are general and can be used to study the response of any linear and isotropic material subjected to EEL and CL spectroscopy.

\begin{figure}[t]
    \centering
    \includegraphics[width=0.8\columnwidth]{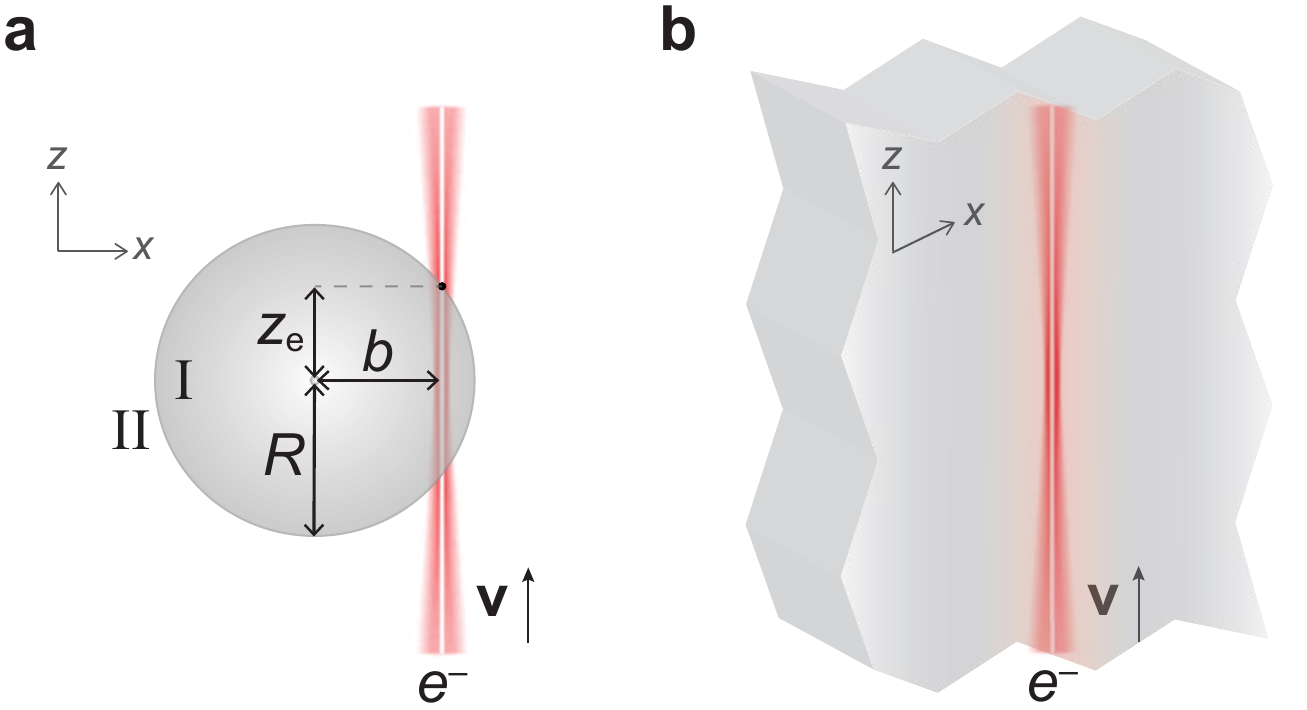}
    \caption{(a) Schematic illustration of the electron beam traveling with velocity $\vect{v} = v\uvec{z}$, and traversing the spherical NP of radius $R$. In the illustration, we assume that the electron travels on the $x$-$z$ plane, entering at point $(x=b, z=-z_e)$ and exiting at $(x=b, z=z_e)$. (b) Schematic illustration of an electron beam of velocity $\vect{v}=v\uvec{z}$ traversing the bulk of an unbound medium of optical parameters $\varepsilon$ and $\mu=1$.}
    \label{fig:SI_illustration}
\end{figure}

We consider the spherical NP of radius $R$ investigated in Section~\ref{sec:results_discussion}, with relative permittivity $\varepsilon$ and relative permeability $\mu=1$. The NP is excited by an electron beam that passes through the bulk of the material, entering at point $(b,\phi_0,-z_e)$ and exiting at $(b,\phi_0,z_e)$ in cylindrical coordinates with respect to the NP center, as illustrated in Fig.~\ref{fig:SI_illustration}a. 
We assume that the recoil of the fast electron in a single scattering event is negligible due to its high velocity, and we therefore work within the non-recoil approximation~\cite{lucas_sunjic}.

To compute the total electromagnetic field arising from the interaction of a fast electron with a spherical NP, we follow the basic steps of Mie theory~\cite{Bohren_Wiley1983}. According to it, the fields inside and surrounding the NP are expanded in terms of spherical waves of unknown expansion coefficients, which are then found by matching the fields at the surface of the NP according to appropriate boundary conditions.
We first express the direct field of the electron $\vect{E}_0$ in terms of spherical waves. Clearly the electromagnetic fields generated by the electron while moving within the NP are different to the ones generated when it travels in the surrounding medium. The electric field $\vect{E}_0^\mathrm{II}$ corresponding to the external trajectory (for $R<r<|\vect{r}_e|$) is given by Eq.~\eqref{E_general_electron}, with modified expansion coefficients
\begin{widetext}
\begin{subequations} \label{exp_coeff_0II}
    \begin{equation} \label{exp_coeff_0IIb}
        b_{\ell m}^{0,\mathrm{II}} = -\frac{ik_0^2 e }{ \varepsilon_0\omega} \frac{m \mathrm{e}^{-im\phi_0}}{\sqrt{\ell(\ell+1)}}  \bigg[\mathcal{M}_{\ell m} K_m \left( \frac{\omega b}{v\gamma_0} \right)  - ik_0  \int_{-z_e}^{z_e} dz \, \mathrm{e}^{i\omega z/v} h_\ell^+(k_0r)  {Y^m_\ell} \left(\theta,0\right)\bigg],  
    \end{equation}
    and
    \begin{align} \label{exp_coeff_0IIa}
       a_{\ell m}^{0,\mathrm{II}}  = \frac{ik_0^2 e}{\varepsilon_0\omega} \frac{\mathrm{e}^{-im\phi_0}}{\sqrt{\ell(\ell+1)}} \bigg[ \frac{\mathcal{N}_{\ell m}}{\beta\gamma_0} & K_m  \left( \frac{\omega b}{v\gamma_0} \right) 
       - \frac{i}{b} \int_{-z_e}^{z_e} dz \,\mathrm{e}^{i\omega z/v} \Big\{ \mathcal{H}_{\ell m}^+(k_0z) + \mathcal{H}_{\ell m}^-(k_0z) \Big\} \bigg],
    \end{align}
\end{subequations}
\end{widetext}
with $z_e = \sqrt{R^2-b^2}$, $r=\sqrt{b^2+z^2}$, and $\theta =\arccos(z/r)$. In Eq.~\eqref{exp_coeff_0IIa} we have set
\begin{align}\label{Flm_coefficients}
    &\mathcal{F}_{\ell m}^\pm (k_nz)= \mp c_\ell^{\pm m}  
    \bigg\{\frac{k_nb^2}{r} f'_\ell (k_nr) Y_\ell^{m \pm 1} (\theta, 0)   \notag \\
    &\pm \frac{zb}{r^2} f_\ell (k_nr)  \left[ c_\ell^{\pm m+ 1} Y_\ell^{m \pm 2} (\theta, 0) - c_\ell^{\pm m} Y_\ell^{m} (\theta, 0) \right] 
      \notag \\
    &+(1\pm m)  f_\ell (k_nr) Y_\ell^{m \pm 1} (\theta, 0) \bigg\}.
\end{align}
that holds for any type of spherical Bessel function $f_\ell (k_n r)$ evaluated in any medium $n$ (the prime here and on any other Bessel function denotes the derivative of the function with respect to the argument). In particular, in Eq.~\eqref{exp_coeff_0IIa} we have  $\mathcal{F}_{\ell m}^\pm (k_nz) = \mathcal{H}_{\ell m}^\pm (k_0z)$ and $f_\ell =h^+_\ell$. 

The electric field generated by the fast electron while traveling inside the NP (for $|\vect{r}_e|<r<R$) acquires the form of outgoing spherical waves, as
\begin{align} \label{E_electron_inside}
    \vect{E}_0^{\mathrm{I}}(\vect{r}, \omega) &= \sum_{\ell =1}^{\infty} \sum_{m=-\ell}^{+\ell}  \Big\{ b_{\ell m}^\mathrm{0,I} h^+_\ell(kr) \vect{X}_{\ell m}(\theta , \phi) \notag  \\
    &\qquad\quad + \frac{i}{k} a_{\ell m}^\mathrm{0,I}   \vecnabla \times h^+_\ell(kr) \vect{X}_{\ell m}(\theta , \phi)  \Big\},
    \end{align}    
where $k=\sqrt{\varepsilon}k_0$ is the wave number inside the NP, and the expansion coefficients are given by
\begin{subequations} \label{exp_coeff_0I}
    \begin{equation}
        b_{\ell m}^{0,\mathrm{I}} = -\frac{ik_0^2 e }{\varepsilon_0\omega} \frac{m\mathrm{e}^{-im\phi_0}}{\sqrt{\ell(\ell+1)}}  ik  \int_{-z_e}^{z_e} dz \,\mathrm{e}^{i\omega z/v} j_\ell(kr)  {Y_\ell^m} \left(\theta,0\right), 
    \end{equation}
    and
    \begin{equation}\label{exp_coeff_0Ia}
        a_{\ell m}^{0,\mathrm{I}}  = \frac{ik_0^2 e}{\varepsilon_0\omega} \frac{\mathrm{e}^{-im\phi_0}}{\sqrt{\ell(\ell+1)}} \frac{i}{b} \int_{-z_e}^{z_e} dz \,\mathrm{e}^{i\omega z/v} \Big\{ \mathcal{J}_{\ell m}^- (kz) + \mathcal{J}_{\ell m}^+ (kz) \Big\}.
    \end{equation}
\end{subequations}
In Eq.~\eqref{exp_coeff_0Ia} we have used expression \eqref{Flm_coefficients} for $\mathcal{F}_{\ell m}^\pm (k_nz) = \mathcal{J}_{\ell m}^\pm (kz)$ and $f_\ell =j_\ell$. 

The total electric field in each region consists of the direct electron field $\vect{E}_0^\mathrm{I/II}$, as well as the fields due to the presence of a boundary (the NP surface) $\vect{E}_\mathrm{B}^\mathrm{I/II}$, and thus is written as $  \vect{E}^\mathrm{I/II}_\mathrm{tot} = \vect{E}_0^\mathrm{I/II} + \vect{E}_\mathrm{B}^\mathrm{I/II}$.
Similar expressions hold for the magnetic field $\vect{H}$, which can be obtained from Faraday's law 
$\vect{H} = \vecnabla \times \vect{E}/(i\omega\mu_0)$. 
Therefore, the total fields inside ($\mathrm{I}$) and surrounding the NP ($\mathrm{II}$), close to the boundary, take the form
\begin{widetext}    
\begin{subequations}\label{fields_expansion_penetrating} 
    \begin{align}
        \vect{E}_\mathrm{tot}^\mathrm{I}(\vect{r}, \omega) =\sum_{\ell =1}^{\infty} \sum_{m=-\ell}^{+\ell}  \Big\{b_{\ell m}^\mathrm{I} j_\ell(kr) \vect{X}_{\ell m}(\theta , \phi) & + \frac{i}{k} a_{\ell m}^\mathrm{I}  \vecnabla \times j_\ell(kr) \vect{X}_{\ell m}(\theta , \phi) \notag \\
        &+b_{\ell m}^{0,\mathrm{I}} h^+_\ell(kr) \vect{X}_{\ell m}(\theta , \phi) + \frac{i}{k} a_{\ell m}^{0,\mathrm{I}}  \vecnabla \times h^+_\ell(kr) \vect{X}_{\ell m}(\theta , \phi) \Big\},
        \label{Efield_expansion_in_penetrating}\\
        \vect{H}_\mathrm{tot}^\mathrm{I}(\vect{r}, \omega) = \frac{1}{Z}\sum_{\ell =1}^{\infty} \sum_{m=-\ell}^{+\ell}   \Big\{ a_{\ell m}^\mathrm{I} j_\ell(kr) \vect{X}_{\ell m}(\theta , \phi) &- \frac{i}{k} b_{\ell m}^\mathrm{I}  \vecnabla \times j_\ell(kr)  \vect{X}_{\ell m}(\theta , \phi)  \notag \\
        &+a_{\ell m}^{0,\mathrm{I}} h^+_\ell(kr) \vect{X}_{\ell m}(\theta , \phi) - \frac{i}{k} b_{\ell m}^{0,\mathrm{I}}  \vecnabla \times h^+_\ell(kr)  \vect{X}_{\ell m}(\theta , \phi)\Big\},
        \label{Hfield_expansion_in_penetrating}
    \end{align}
    \begin{align}
        \vect{E}_\mathrm{tot}^\mathrm{II}(\vect{r}, \omega) = \sum_{\ell =1}^{\infty} \sum_{m=-\ell}^{+\ell}  \Big\{ b_{\ell m}^\mathrm{II} h^+_\ell(k_0r) \vect{X}_{\ell m}(\theta , \phi) &+ \frac{i}{k_0} a_{\ell m}^\mathrm{II}  \vecnabla \times h^+_\ell(k_0r) \vect{X}_{\ell m}(\theta , \phi)   \notag \\      
        &+ b_{\ell m}^{0,\mathrm{II}} j_\ell(k_0r) \vect{X}_{\ell m}(\theta , \phi) + \frac{i}{k_0} a_{\ell m}^{0,\mathrm{II}}  \vecnabla \times j_\ell(k_0r) \vect{X}_{\ell m}(\theta , \phi) \Big\}, 
        \label{Efield_expansion_out_penetrating}\\
        \vect{H}_\mathrm{tot}^\mathrm{II}(\vect{r}, \omega)= \frac{1}{Z_0}\sum_{\ell =1}^{\infty} \sum_{m=-\ell}^{+\ell}   \Big\{ a_{\ell m}^\mathrm{II} h_\ell^+(k_0r) \vect{X}_{\ell m}(\theta , \phi) &- \frac{i}{k_0} b_{\ell m}^\mathrm{II}  \vecnabla \times h_\ell^+(k_0r) \vect{X}_{\ell m}(\theta , \phi) \notag \\
        &+a_{\ell m}^{0,\mathrm{II}}   j_\ell(k_0r) \vect{X}_{\ell m}(\theta , \phi) - \frac{i}{k_0} b_{\ell m}^{0,\mathrm{II}}  \vecnabla \times j_\ell(k_0r) \vect{X}_{\ell m}(\theta , \phi) \Big\},
        \label{Hfield_expansion_out_penetrating}
    \end{align}
\end{subequations}
\end{widetext}
where $Z = \sqrt{\mu_0/(\varepsilon\varepsilon_0)}$ is the impedance inside the NP. 

At the interface $r=R$, the total fields must satisfy the boundary conditions that require the continuity of the tangential components of the $\vect{E}$ and $\vect{H}$ field~\cite{jackson}, leading to the following set of equations
\begin{subequations}\label{mie_bc}
    \begin{align}
        \vect{X}_{\ell m}^* \cdot \vect{E}^\mathrm{I}_\mathrm{tot} \Big|_{r=R} &= 
        \vect{X}_{\ell m}^* \cdot \vect{E}^\mathrm{II}_\mathrm{tot} \Big|_{r=R}, \\
         \vect{X}_{\ell m}^* \cdot \vect{H}^\mathrm{I}_\mathrm{tot}\Big|_{r=R} &= 
         \vect{X}_{\ell m}^*\cdot \vect{H}^\mathrm{II}_\mathrm{tot} \Big|_{r=R}, \\
        \left[ \uvec{r} \times \vect{X}_{\ell m}^* \right] \cdot \vect{E}^\mathrm{I}_\mathrm{tot} \Big|_{r=R} &= \left[ \uvec{r} \times \vect{X}_{\ell m}^* \right] \cdot \vect{E}^\mathrm{II}_\mathrm{tot} \Big|_{r=R}, \\  
        \left[ \uvec{r} \times \vect{X}_{\ell m}^* \right] \cdot \vect{H}^\mathrm{I}_\mathrm{tot} \Big|_{r=R} &= \left[ \uvec{r} \times \vect{X}_{\ell m}^*\right] \cdot \vect{H}^\mathrm{II}_\mathrm{tot} \Big|_{r=R}.
    \end{align}     
\end{subequations}
Using the orthonormality relations of the vector spherical harmonics~\cite{jackson}, we obtain the unknown expansion coefficients
\begin{subequations}
\label{expansion_coefficients_penetrating}
    \begin{align}
    a_{\ell m}^\mathrm{I} &= T_{\mathrm{E}\ell}^{11} a_{\ell m}^{0,\mathrm{I}} + T_{\mathrm{E}\ell}^{21} a_{\ell m}^{0,\mathrm{II}}, \\
    a_{\ell m}^\mathrm{II} &= T_{\mathrm{E}\ell}^{12} a_{\ell m}^{0,\mathrm{I}} + T_{\mathrm{E}\ell}^{22} a_{\ell m}^{0,\mathrm{II}},  \\
    b_{\ell m}^\mathrm{I} &= T_{\mathrm{M}\ell}^{11} b_{\ell m}^{0,\mathrm{I}} + T_{\mathrm{M}\ell}^{21} b_{\ell m}^{0,\mathrm{II}}, \\
    b_{\ell m}^\mathrm{II} &= T_{\mathrm{M}\ell}^{12} b_{\ell m}^{0,\mathrm{I}} + T_{\mathrm{M}\ell}^{22} b_{\ell m}^{0,\mathrm{II}}.
    \end{align}
\end{subequations}
Adopting the notation of the Riccati--Bessel functions $\Psi_\ell (x) = xj_\ell(x)$ and $\xi_\ell (x) = xh^+_\ell(x)$, the Mie coefficients entering Eqs.~\eqref{expansion_coefficients_penetrating} are given by
\begin{subequations} \label{Mie_coefficients}
    \begin{align}
        T_{\mathrm{E}\ell}^{22} &=  \frac{ \varepsilon j_\ell(  kR) \Psi_\ell' (  k_0R) -   \Psi_\ell'(  kR) j_\ell (  k_0R)}{  h^+_\ell(  k_0R) \Psi_\ell' (  kR) -  \varepsilon \xi_\ell'(  k_0R) j_\ell (  kR)},\\
        T_{\mathrm{M}\ell}^{22} &=\frac{ j_\ell(  kR) \Psi_\ell' (  k_0R) -   \Psi_\ell'(  kR) j_\ell (  k_0R)}{  h^+_\ell(  k_0R) \Psi_\ell' (  kR) -  \xi_\ell'(  k_0R) j_\ell (  kR)}, \\  
        T_{\mathrm{E}\ell}^{21} &= -\frac{i\sqrt{\varepsilon}/(k_0R)}{h_\ell^+(  k_0R)\Psi'_\ell (  kR) - \varepsilon\xi'_\ell (  k_0R)j_\ell (  kR) }, \\
        T_{\mathrm{M}\ell}^{21} &=- \frac{i/(k_0R)}{h_\ell^+(  k_0R)\Psi'_\ell (  kR) - \xi'_\ell (  k_0R)j_\ell (  kR) }, \\
         T_{\mathrm{E}\ell}^{11} &=  \frac{\varepsilon\xi'_\ell (  k_0R)h_\ell^+(  kR) - h_\ell^+(  k_0R)\xi'_\ell (  kR)}{h_\ell^+(  k_0R)\Psi'_\ell (  kR) - \varepsilon\xi'_\ell (  k_0R)j_\ell (  kR) }, \\
         T_{\mathrm{M}\ell}^{11} &= \frac{\xi'_\ell (  k_0R)h_\ell^+(  kR) - h_\ell^+(  k_0R)\xi'_\ell (  kR)}{h_\ell^+(  k_0R)\Psi'_\ell (  kR) - \xi'_\ell (  k_0R)j_\ell (  kR) },  \\ 
        T_{\mathrm{E}\ell}^{12} &= -\frac{i/(k_0R)}{h_\ell^+(  k_0R)\Psi'_\ell (  kR) - \varepsilon\xi'_\ell (  k_0R)j_\ell (  kR) }, \\
        T_{\mathrm{M}\ell}^{12} &=    -\frac{i/(\sqrt{\varepsilon}k_0R)}{h_\ell^+(  k_0R)\Psi'_\ell (  kR) - \xi'_\ell (  k_0R)j_\ell (  kR) }.
    \end{align}
\end{subequations}

We now have all ingredients required to evaluate the fields of Eqs.~\eqref{fields_expansion_penetrating}, and calculate the EEL an CL probability.

\section{EEL and CL probability}
\label{sec:appedix_d}

\subsection*{EEL Probability}
\label{subsec:appedix_d1}

The fast electron loses part of its kinetic energy due to 
energy transfer to the optical modes sustained in the structure; here, the plasmon oscillations sustained in metals. The probability of the electron losing energy $\hbar\omega$ is related to the work done by the electron moving against the induced field along the electron trajectory~\cite{GarciadeAbajo:1999prb}, and is given by Eq.~\eqref{EELS_general}, or reshaped as
\begin{equation} \label{G_EELS_general}
    \Gamma_\mathrm{EEL} (\omega) 
    = \frac{e}{\pi\hbar\omega} \int_{-\infty}^{+\infty} dz\, \mathrm{Re} \big\{ \mathrm{e}^{-i\omega z/v} \,\uvec{z} \cdot \vect{E}_\mathrm{ind} (\vect{r}_e, \omega) \big\}.
\end{equation}
The induced field can be found as the total field minus the electron field $\vect{E}_0^\mathrm{air}$ in the absence of the structure, i.e. $\vect{E}_\mathrm{ind} = \vect{E}_\mathrm{tot} -\vect{E}_0^\mathrm{air}$. Eventually, in the two regions the induced field is
$\vect{E}_\mathrm{ind}^\mathrm{II} = \vect{E}_\mathrm{B}^\mathrm{II}$, and 
$\vect{E}_\mathrm{ind}^\mathrm{I} = \vect{E}_0^\mathrm{I} + \vect{E}_\mathrm{B}^\mathrm{I}-\vect{E}_0^\mathrm{air}$
and Eq.~\eqref{G_EELS_general} yields
\begin{widetext}
\begin{align} 
    \Gamma_\mathrm{EEL} (\omega)
    =\frac{e}{\pi\hbar\omega} \mathrm{Re} \bigg[   \int_{-\infty}^{+\infty} dz\,\mathrm{e}^{-i\omega z/v}\, \uvec{z} \cdot \vect{E}_\mathrm{ind}^\mathrm{II} (\vect{r}_e, \omega) 
    - \int_{-z_e}^{z_e} dz\, \mathrm{e}^{-i\omega z/v}\, \uvec{z} \cdot \vect{E}_\mathrm{ind}^\mathrm{II} (\vect{r}_e, \omega)
    +   \int_{-z_e}^{z_e} dz \, \mathrm{e}^{-i\omega z/v}\, \uvec{z} \cdot \vect{E}_\mathrm{ind}^\mathrm{I} (\vect{r}_e, \omega)  \bigg] , 
\end{align}
where we have split the integral running along the electron trajectory, into the path of the electron within and outside the NP. 
We further decompose the EEL probability into three terms as follows

\begin{align}
    \Gamma_\mathrm{EEL} (\omega)  
    =\frac{e}{\pi\hbar\omega} \mathrm{Re}\bigg[& \underbrace{ \int_{-\infty}^{+\infty} dz  \, \mathrm{e}^{-i\omega z/v} \,\uvec{z} \cdot \vect{E}_\mathrm{B}^\mathrm{II} (\vect{r}_e, \omega) - \int_{-z_e}^{z_e} dz\, \mathrm{e}^{-i\omega z/v} \,\uvec{z} \cdot \vect{E}_\mathrm{B}^\mathrm{II} (\vect{r}_e, \omega) }_{\Gamma_\mathrm{surf}}  
     + \underbrace{ \int_{-z_e}^{z_e} dz\, \mathrm{e}^{-i\omega z/v} \,\uvec{z} \cdot  \vect{E}_\mathrm{B}^\mathrm{I} (\vect{r}_e, \omega) }_{\Gamma_\mathrm{Begr}} \notag \\
    &+\underbrace{ \int_{-z_e}^{z_e} dz\, \mathrm{e}^{-i\omega z/v} \,\uvec{z} \cdot  \big\{\vect{E}_0^\mathrm{I} (\vect{r}_e, \omega)  -\vect{E}_0^\mathrm{air}  (\vect{r}_e, \omega) \big\}}_{\Gamma_\mathrm{bulk}} \bigg]
    = \Gamma_\mathrm{surf} + \Gamma_\mathrm{Begr} +  \Gamma_\mathrm{bulk}.
    \label{EELS_decomp_step}    
\end{align}   
\end{widetext}
In the decomposition presented in Eq.~\eqref{EELS_decomp_step}, the $\Gamma_\mathrm{bulk}$ term corresponds to the energy lost to the excitation of bulk modes of the unbound medium modified and reduced by the \emph{Begrenzung} term $\Gamma_\mathrm{Begr}$, that accounts for the presence of a boundary. 
The remaining terms are grouped together and referred to as the \textquote{surface term} $\Gamma_\mathrm{surf}$, to indicate that they only contain contributions to the EEL probability pertaining to the excitation of LSPs (see Fig.~\ref{fig:analytics}c). We stress, however, that said name serves classification purposes here, and it might not be appropriate in the study of non-plasmonic materials that support modes of different characteristics. For high-index dielectric nanospheres, for instance, that host Mie resonances in their volume, the description \textquote{surface} is not appropriate. 

To derive the bulk contribution to EELS, we make use of Eq.~\eqref{eq:field0_z_comp}, which determines the field generated by the fast electron traveling distance $2z_e$ within an infinite medium of relative permittivity $\varepsilon$ and relative permeability $\mu=1$. The corresponding EEL probability $\Gamma_\mathrm{bulk} (\omega)$
can be found via the momentum-resolved EEL probability $P_\mathrm{EEL} (q, \omega) $
 \begin{align}
    P_\mathrm{EEL} (q, \omega) = -\frac{e^2qz_e}{\pi^2\hbar \varepsilon_0 v^2}& \mathrm{Im} \bigg\{ \frac{1}{[q^2-k^2+ (\omega/v)^2]\gamma^2 \varepsilon} \notag \\
    - &\frac{1}{[q^2-k_0^2+ (\omega/v)^2]\gamma_0^2} \bigg\}.
\end{align}   
Then the EEL probability $\Gamma_\mathrm{EEL}(\omega)$ is obtained by 
\begin{align} \label{eq:integral_bulk}
    \Gamma_\mathrm{EEL}(\omega) = \int_0^{q_\mathrm{c}} dq \, P_\mathrm{EEL} (q, \omega), 
\end{align}
leading to Eq.~\eqref{bulk_term}. The isolation of the bulk term inevitably leads to the introduction of the free parameter $q_{\mathrm{c}}$; considering that the electron transfers transverse (with respect to the electron trajectory) momentum $q$ upon losing energy $\hbar\omega$, we impose a cutoff $q_{\mathrm{c}}$ to the maximum transverse momentum collected, in order to ensure finiteness of the EEL probability~\cite{lucas_sunjic}. Its value can be determined by the half-aperture collection angle of the microscope spectrometer, according to Eq.~\eqref{Eq:cutoff}.

Inserting the fields of Eqs.~\eqref{fields_expansion_penetrating} in Eq.~\eqref{EELS_decomp_step} we obtain
\begin{equation} \label{EELS_decomp2}
    \Gamma_\mathrm{EEL} = \Gamma_\mathrm{surf}^\mathrm{M}+ \Gamma_\mathrm{surf}^\mathrm{E} + \Gamma_\mathrm{Begr}^\mathrm{M} + \Gamma_\mathrm{Begr}^\mathrm{E} +  \Gamma_\mathrm{bulk},  
\end{equation}
where we have further decomposed the surface and Begrenzung terms into contributions from electric (E) and magnetic-type (M) multipoles, that take the form
\begin{widetext}
\begin{subequations}\label{EELS_final}
\begin{gather} 
    \Gamma_\mathrm{surf}^{\mathrm{M}}(\omega) 
    = \frac{e}{\pi\hbar\omega}\mathrm{Re} \sum_{\ell =1}^{\infty} \sum_{m=-\ell}^{+\ell} \mathrm{e}^{im\phi_0}\frac{mb_{\ell  m}^\mathrm{II}}{\sqrt{\ell (\ell +1)}}   \bigg[\frac{\mathcal{M}_{\ell m}^*}{ik_0 } K_m \left( \frac{\omega b}{v\gamma_0} \right) 
   -\int_{-z_e}^{z_e} dz \, \mathrm{e}^{-i\omega z/v} h_\ell^+(k_0r)  {Y^m_\ell} \left(\theta,0\right)\bigg], \\
   \Gamma_\mathrm{surf}^{\mathrm{E}}(\omega)      
     = -\frac{e}{\pi\hbar\omega} \mathrm{Re}  \sum_{\ell =1}^{\infty} \sum_{m=-\ell}^{+\ell} \mathrm{e}^{im\phi_0} \frac{ a_{\ell  m}^\mathrm{II}} {\sqrt{\ell (\ell +1)}}  \bigg[  \frac{{\mathcal{N}_{\ell m}}^*}{ik_0\beta\gamma_0} K_m \left( \frac{\omega b}{v\gamma_0} \right) 
     - \frac{1}{k_0b}\int_{-z_e}^{z_e} dz \,\mathrm{e}^{-i\omega z/v} \big\{ \mathcal{H}_{\ell m}^+(k_0z) + \mathcal{H}_{\ell m}^-(k_0z) \big\}\bigg], \\
    \Gamma_\mathrm{Begr}^{\mathrm{M}}  (\omega) 
      =\frac{e}{\pi\hbar\omega}\mathrm{Re} \sum_{\ell =1}^{\infty} \sum_{m=-\ell}^{+\ell} \mathrm{e}^{im\phi_0}\frac{mb_{\ell  m}^\mathrm{I}}{\sqrt{\ell (\ell +1)}}  \int_{-z_e}^{z_e} dz \, \mathrm{e}^{-i\omega z/v} j_\ell(kr)  {Y^m_\ell} \big(\theta,0\big), \\
    \Gamma_\mathrm{Begr}^{\mathrm{E}}  (\omega) =
    - \frac{e}{\pi\hbar\omega} \mathrm{Re}  \sum_{\ell =1}^{\infty} \sum_{m=-\ell}^{+\ell} \mathrm{e}^{im\phi_0} \frac{ a_{\ell  m}^\mathrm{I}} {\sqrt{\ell (\ell +1)}} \frac{1}{kb}\int_{-z_e}^{z_e} dz \,\mathrm{e}^{-i\omega z/v} \big\{ \mathcal{J}_{\ell m}^+(kz) + \mathcal{J}_{\ell m}^-(kz) \big\}, 
\end{gather}
\end{subequations}
\end{widetext}
making again use of expression \eqref{Flm_coefficients}.
We note here that the exponential term $\mathrm{e}^{im\phi_0}$ in the formulas above need not be taken into account, since it is canceled out by the term $\mathrm{e}^{-im\phi_0}$ included in the expansion coefficients $a/b_{\ell m}^{0,\textrm{I/II}} $ of Eqs.~\eqref{exp_coeff_0II}, \eqref{exp_coeff_0I}. 

To avoid factoring-in the direct field of the electron, we correct the expansion coefficients of Eqs.~\eqref{expansion_coefficients_penetrating} as follows
\begin{subequations}\label{exp_coeff_corrected}
    \begin{align}
    a_{\ell m}^\textrm{I} &= T_{\mathrm{E}\ell}^{11} a_{\ell m}^{0,\textrm{I}} + T_{\mathrm{E}\ell}^{21} a_{\ell m}^{0,\textrm{II}} - a_{\ell m}^{0,\textrm{II}},\\
    a_{\ell m}^\textrm{II} &= T_{\mathrm{E}\ell}^{12} a_{\ell m}^{0,\textrm{I}} + T_{\mathrm{E}\ell}^{22} a_{\ell m}^{0,\textrm{II}} - a_{\ell m}^{0,\textrm{I}}|_\mathrm{air},\\
    b_{\ell m}^\textrm{I} &= T_{\mathrm{M}\ell}^{11} b_{\ell m}^{0,\textrm{I}} + T_{\mathrm{M}\ell}^{21} b_{\ell m}^{0,\textrm{II}}-  b_{\ell m}^{0,\textrm{II}}, \\
    b_{\ell m}^\textrm{II} &= T_{\mathrm{M}\ell}^{12} b_{\ell m}^{0,\textrm{I}} + T_{\mathrm{M}\ell}^{22} b_{\ell m}^{0,\textrm{II}}- b_{\ell m}^{0,\textrm{I}}|_\mathrm{air},
    \end{align}
\end{subequations}
where the notation $a/b_{\ell m}^{0,\textrm{I}}|_\mathrm{air}$ indicates evaluation of the terms given by Eq.~\eqref{exp_coeff_0I} in air ($\varepsilon=1$, $k=k_0$). This correction ensures a zero probability in the absence of the structure.

For an aloof electron trajectory, we set $z_e=0$ and find the expressions presented in Refs.~\cite{GarciadeAbajo:1999prb,GarciadeAbajo:2010rmp}.

\subsection*{CL probability}
\label{subsec:appedix_d2}

Upon coupling to the propagating electron, the radiative excited modes decay by emitting radiation to the far field. The probability $\Gamma_\textrm{CL}$ of emitting a photon of energy $\hbar\omega$ is obtained via the Poynting flux of Eq.~\eqref{CL}. By inserting the induced field $\vect{E}_\mathrm{ind}^\mathrm{II} = \vect{E}_\mathrm{B}^\mathrm{II}$, and letting $r \to \infty$, we obtain
\begin{equation} \label{CL_final}
       \Gamma_\textrm{CL} (\omega) =\frac{1}{\pi \hbar\omega Z_0k^2_0}  \sum_{\ell =1}^{\infty} \sum_{m=-\ell}^{+\ell}  \left[\left| b_{\ell  m}^\textrm{II} \right|^2 + \left| a_{\ell  m}^\textrm{II} \right|^2   \right].  
\end{equation}
As in the EELS calculation, we remove the contribution of the direct electron field by modifying coefficients $a_{\ell  m}^\textrm{II}$ and $b_{\ell  m}^\textrm{II}$ according to Eqs.~\eqref{exp_coeff_corrected}. The final result agrees with the expression for $\Gamma_\textrm{CL}$ presented in Ref.~\cite{matsukata:2021}.

As a final comment, we mention that the EEL and CL probabilities in Eqs.~\eqref{eq:integral_bulk}, \eqref{EELS_final} and \eqref{CL_final} are given in units of seconds (s), whereas in all relevant figures in the main text they are presented in units of inverse energy (eV)$^{-1}$. To perform the unit conversion one can simply divide by the reduced Planck constant $\hbar$.

\begin{figure}[b]
    \centering
    \includegraphics[width=\columnwidth]{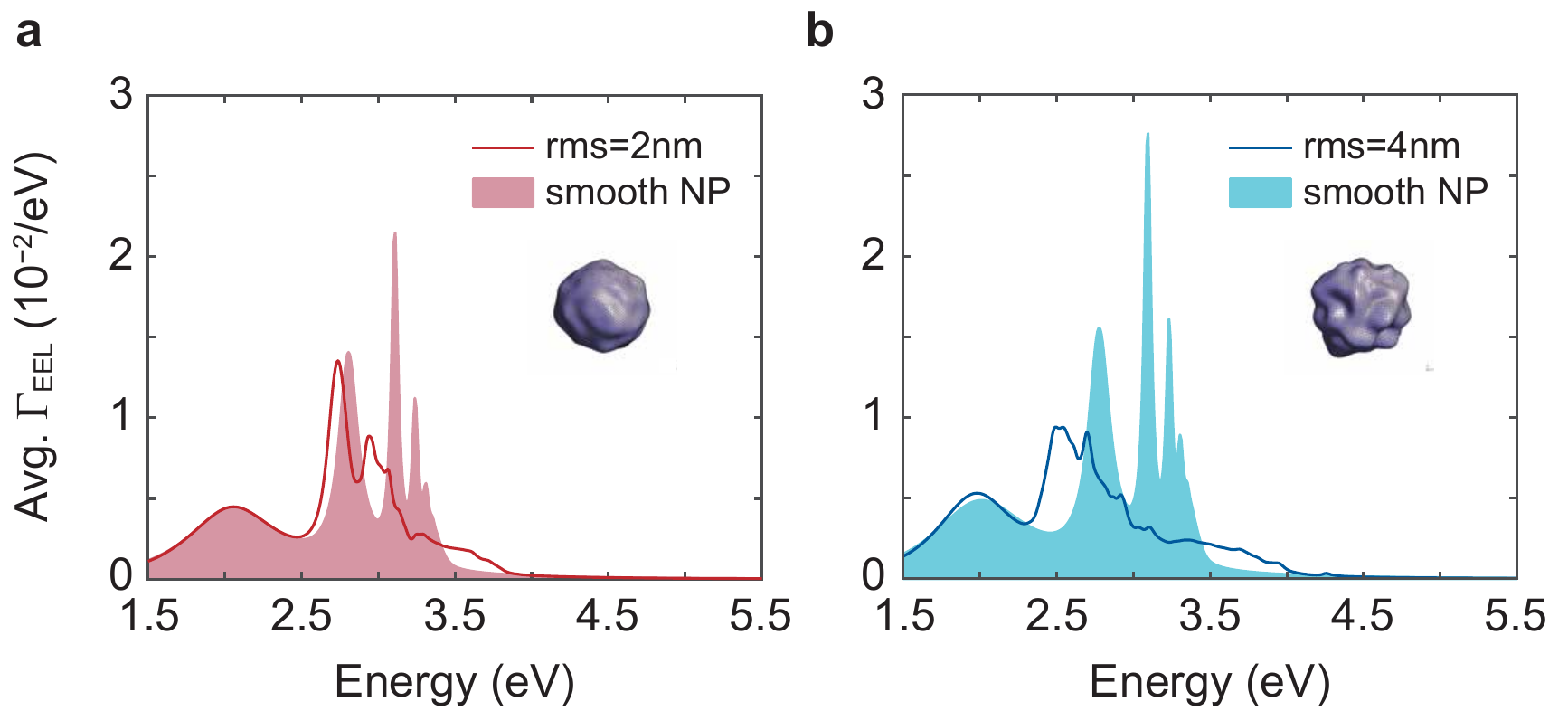}
    \caption{Average EEL probability in the interaction between a spherical NP featuring surface roughness and an electron beam passing with velocity $v = 0.7 c$ (kinetic energy $\approx 200$\,keV) at distance $b=100$\,nm. The solid lines correspond to NPs, whose shapes deviate from that of a perfectly smooth sphere of radius $R=75$\,nm  by (a) $\mathrm{rms}=2$\,nm (dark red curve) and (b) $\mathrm{rms}=4$\,nm (dark blue curve). The pink (panel a) and cyan areas (panel b) show the analytically derived EEL probability for a smooth sphere of effective radius (a) $R_\mathrm{eff}=76.2$\,nm, and (b) $R_\mathrm{eff}=79.6$\,nm.}
    \label{fig:rough_sphere_comparison}
\end{figure}

\section{Corrugated nanospheres compared to smooth ones of larger size}
\label{sec:appendix_e}

In Fig.~\ref{fig:rough_sphere}, we ascribe the redshift observed in the average EEL spectra of NPs with surface roughness, as compared to the smooth spheres, to the increase of the area on the NP surface. 
Figs.~\ref{fig:rough_sphere_comparison}a and b present the average EEL probability of two NPs with $\mathrm{rms}=2$\,nm (dark red curve), and $\mathrm{rms}=4$\,nm (dark blue curve), respectively (same data as the ones shown in Fig.~\ref{fig:rough_sphere} with the same color). By evaluating the area from the mesh parameters we obtain an effective radius of $R_\mathrm{eff}=76.2$\,nm for $\mathrm{rms}=2$\,nm, and $R_\mathrm{eff}=79.6$\,nm for $\mathrm{rms}=4$\,nm. In order to confirm our argument, in Fig.~\ref{fig:rough_sphere_comparison} we juxtapose the EEL probability calculated analytically for a smooth NP of radius equal to the calculated $R_\mathrm{eff}$ in each case. Indeed, the redshift of the lower-energy peak (electric dipolar mode) is very well reproduced by smooth NPs of the same area. Due to the symmetry breaking, higher-order modes exhibit a more complex behavior; additionally to the shift, they also feature energy splits.

\hfill

\bibliography{references}

\end{document}